\documentstyle[preprint,eqsecnum,aps]{revtex}

\begin{document}

\newcommand{\beq}{\begin{equation}}
\newcommand{\eeq}{\end{equation}}
\newcommand{\beqa}{\begin{eqnarray}}
\newcommand{\eeqa}{\end{eqnarray}}
\newcommand{\boldsigma}{\mbox{\boldmath$\sigma$}}
\newcommand{\boldtau}{\mbox{\boldmath$\tau$}}
\newcommand{\ott}{\boldtau_i \cdot \boldtau_j}
\newcommand{\oss}{\boldsigma_i\cdot\boldsigma_j}
\newcommand{\os}{S_{ij}}
\newcommand{\ols}{({\bf L\cdot S})_{ij}}
\newcommand{\hovm}{\frac{\hbar^2}{m}}
\draft

\title{Spin-Isospin Structure and Pion Condensation
in Nucleon Matter}

\author{A. Akmal\cite{aa} and V. R. Pandharipande\cite{vrp}}
\address{Department of Physics, University of Illinois at Urbana-Champaign,
         1110 W. Green St., \\
Urbana, IL 61801}

\date{\today}
\maketitle
\begin{abstract}

We report variational calculations of symmetric nuclear matter 
and pure neutron matter, using the new Argonne $v_{18}$ 
two-nucleon and Urbana IX three-nucleon interactions. 
At the equilibrium density of 0.16 fm$^{-3}$ the
two-nucleon densities in symmetric nuclear matter 
exhibit a short-range spin-isospin structure similar to
that found in light nuclei. We
also find that both symmetric nuclear matter and
pure neutron matter undergo transitions to phases with pion
condensation at densities of 0.32 fm$^{-3}$ and 0.2 fm$^{-3}$, respectively.
Neither transtion occurs with the Urbana $v_{14}$ two-nucleon
interaction, while only the transition in neutron matter occurs
with the Argonne $v_{14}$ two-nucleon interaction.
The three-nucleon interaction is required for the transition
to occur in symmetric nuclear matter, whereas the transition
in pure neutron matter occurs even in its absence.
The behavior of the isovector spin-longitudinal response and the
pion excess in the vicinity of the transition, and the
model dependence of the transition are discussed.

\end{abstract}

\newpage
\section{INTRODUCTION}

A central problem in nuclear many-body theory is the
prediction from realistic models of nuclear forces,
of properties of uniform nuclear matter in
stable nuclei and in neutron stars. 
This problem has been the focus of much attention
since the pioneering work of
Brueckner, Levinson and Mahmoud \cite{blm} in 1954. Early
work in this field, based primarily on Brueckner-Bethe-Goldstone
many-body theory, was reviewed by Bethe \cite{bethe} in 1971.
A variational method based on hypernetted chain summation
techniques (VCS, previously denoted FHNC-SOC) was developed
in the 1970s \cite{pw}, particularly to include
the effects of many-body correlations, presumably important in
dense neutron star matter. Calculations performed since then have
confirmed that many-body clusters make significant contributions
to the binding energies of equilibrium nuclear matter
and  light nuclei \cite{pw,day,dw,pwp,forest,pscppr}.

Friedman and Pandharipande \cite{fp} (hereafter denoted FP) carried out
detailed calculations of the equation of state of symmetric
nuclear matter (SNM) containing equal numbers of protons and
neutrons and of pure neutron matter (PNM) with VCS, using the
Urbana $v_{14}$ model of the NN interaction \cite{lp2}. In these
calculations, effects of many-body forces were introduced partly
via a small density dependence in the NN interaction and partly
by adding a density-dependent contribution, attributed to
the attractive two-pion exchange three-nucleon interaction \cite{lp3}. 
The parameters of these contributions were 
adjusted to reproduce the empirical properties of SNM, and
methods were developed to interpolate between SNM and PNM
\cite{lp4}. The main weakness in this approach is that the
effect of many-body interactions on the structure of matter
is ignored.

In the 1980s, a variational theory using Monte Carlo methods
(VMC) was developed for light nuclei \cite{lps}. Correlation
operators with the same structure were employed in both
VMC and VCS methods, with the aim of obtaining a unified
description of light nuclei and nuclear matter.
The Urbana VII (UVII) model of three nucleon interactions, containing
the Fujita-Miyazawa two-pion exchange three-nucleon interaction
\cite{fm} and a phenomenological, shorter range interaction,
was added to the Hamiltonian. The parameters of the UVII were
determined by reproducing the binding energies of $^3$H, $^4$He
and the equilibrium density of SNM using approximate VMC and
VCS calculations \cite{spw}.

Wiringa, Fiks and Fabrocini \cite{wff} (hereafter denoted WFF)
calculated the equation of state of SNM and PNM with an improved
VCS method using the Urbana and Argonne~\cite{wsa} $v_{14}$
models of the NN interaction and the Urbana VII three-nucleon
interaction. 
Their calculations of PNM indicated a transition to a new phase, 
possibly including pion condensation,
at $\rho\sim 0.24$ fm$^{-3}$ with the Argonne~$v_{14}$,
though not with Urbana~$v_{14}$.
No evidence of a similar transition in SNM was found by WFF.

Since the pioneering work of Migdal \cite{migdal} and of
Sawyer and Scallapino \cite{ss}, many investigators have 
used effective interactions to study
the possibility of pion condensation in SNM and PNM.
These efforts were recently reviewed by
Kunihiro {\it et al.} \cite{kun}. In these studies, the $\Delta$-resonance
is explicitly considered as a non-nucleonic degree of freedom.
In contrast, WFF considered only nucleonic degrees of freedom and absorbed the 
effect of the $\Delta$-resonance into the two- and three-nucleon
interactions. The approaches used by Migdal and by WFF to
study pion condensation are thus very different. Although the effect
of this type of transition on the equation of state is 
relatively small, it can have important consequences for
the cooling and evolution of neutron stars \cite{pethick}.

In the late 1980s Carlson \cite{carlson88} developed the
Green's function Monte Carlo (GFMC) method, with which exact
calculations of light nuclei are possible, starting from  
Hamiltonians with realistic two- and three-nucleon interactions.
The GFMC method can be used to determine some of the parameters of
the three-nucleon interaction from exact calculations of
nuclear binding energies.

Recently, Wiringa, Stoks and Schiavilla \cite{wss} made significant
improvements to the Urbana-Argonne $v_{14}$ models of the NN 
interaction by including isospin symmetry breaking terms.
Their resulting Argonne $v_{18}$ is one of the new models
fit to the Nijmegen NN scattering database \cite{nij1,nij2}. 
Using this interaction, along with the GFMC method, Pudliner {\it et al.} 
\cite{pud1} obtained parameters for the Urbana IX (UIX) model
of the three-nucleon interaction. Exact GFMC calculations by
Pudliner {\it et al.} \cite{pud2,pud3} have shown that the 
Argonne $v_{18}$ and
Urbana IX interactions provide a good description of all bound
states, up to seven nucleons. While all observed energies are
not exactly reproduced, in each case the difference between 
theory and experiment is much smaller than the contribution
of the three-nucleon interaction.

In the present work, we employ the Argonne $v_{18}$ and Urbana IX
interactions to study the properties of SNM and PNM, using the
VCS method. We have introduced further improvements to the
method, in order to address the relatively strong momentum dependence of the
Argonne $v_{18}$ interaction. These improvements are described
in Appendices A and B.

With these interactions, we find that both SNM and PNM undergo
transitions to a new phase at densities of $\sim 0.32$ fm$^{-3}$
and $\sim 0.2$ fm$^{-3}$, respectively. The transitions are
similar in nature to that found by WFF in PNM, using the
Argonne $v_{14}$ and Urbana model VII.

In this paper, we discuss the phase transition and the spin-isospin
structure of normal SNM at equilibrium density. Additional results for
the equation of state of dense matter, including relativistic effects,
will be reported separately. The plan of the present paper is as follows.
The Hamiltonian and the VCS method
are reviewed in Sections II and III. In Section IV, we present
various pair distribution functions of interest, at both equilibrium and
transition densities. These results clearly indicate that in normal SNM at
$\rho =0.16$ fm$^{-3}$, the tensor correlations in T=0 states
have near maximum possible strength at $r\sim 1$ fm. The results also
indicate the important  role played by the short-range parts of $\oss\ott$ and
$\oss$ interactions in determining the transition density, as 
predicted by Migdal. In Migdal's
approach, the strength of the short-range part of the $\oss\ott$ interaction
is represented by the Landau-Migdal parameter $g^\prime$. We also comment
on the model dependence of this transition in Section IV, by comparing
the Urbana $v_{14}$ and the Argonne $v_{14}$ and $v_{18}$ interactions.
Results for the sums of isovector spin-longditudinal responses
of SNM and PNM are presented in Section V.
The corresponding results for the strengths of pion exchange interactions
and pion fields appear in section VI.
These results clearly indicate that the high density phase includes
pion condensation, as anticipated by WFF. Results for SNM are
presented at normal density and at the transition density.
The response of SNM at normal density exhibits little indication
of the phase transition, which is evidently of first order in 
the present approach.  Conclusions are presented in Section VII.

\section{THE HAMILTONIAN}
The Argonne $v_{18}$ two-nucleon interaction has the form:
\begin{equation}
v_{18,ij}=\sum_{p=1,18}v^p(r_{ij})O^p_{ij}+v_{em} .
\end{equation}
The electromagnetic part $v_{em}$ consists of Coulomb and
magnetic interactions in the nn, np and pp pairs, and it is omitted
from all nuclear matter studies. The strong interaction
part of the potential includes fourteen isoscalar terms with operators:
\begin{equation}
O^{p=1,14}_{ij}   =
\left[
1,\oss, \os,\ols, L^2, L^2\oss, \ols^2
\right] \otimes \left[ 1,\ott \right].
\end{equation}
By convention, operators with even  $p\leq$ 14 include the $\ott$
factor, while those with odd  $p\leq$ 13 are independent of isospin.
The operators with $p=$15, 16 and 17 are isotensor:
\begin{equation}
O^{p=15,17}_{ij}  =
(3\tau_{zi}\tau_{zj}-\ott)\otimes
(1,\oss,S_{ij}).
\end{equation}
The interactions associated with operators $p=16$ and 17 are generated
by the mass difference between neutral and charged pions, while the
phenomenological interaction with the $p=$15 operator is necessary to
fit the difference between np and pp, $^1S_0$ phase shifts. Finally,
the isovector interaction with the operator,
\begin{equation}
O^{18}_{ij} = (\tau_{zi}+\tau_{zj}),
\end{equation}
is required to fit the nn scattering length and effective range along with
the np and pp data. The interaction constructed with the above elements
provides an accurate fit to the NN scattering data in the Nijmegen database,
using the kinetic energy operator:
\begin{equation}
K=\sum_i -\frac{\hbar^2}{4}\left[ \left( \frac{1}{m_p} + \frac{1}{m_n} \right)
+ \left( \frac{1}{m_p} - \frac{1}{m_n} \right) \tau_{zi} \right] \nabla_i^2,
\end{equation}
which takes into account the mass difference between protons and neutrons.

The Urbana models of $V_{ijk}$ contain two isoscalar terms:
\begin{equation}
V_{ijk}=V^{2\pi}_{ijk} + V^{R}_{ijk} \ .
\end{equation}
The first term represents the Fujita-Miyazawa two-pion exchange interaction:
\beqa
V^{2\pi}_{ijk}=&\sum_{cyc} A_{2\pi} \left( \left\{\ott,
\boldtau_i \cdot \boldtau_k \right\}
\left\{ X_{ij},X_{ik} \right\}
+ \frac{1}{4} [\ott, \boldtau_i \cdot \boldtau_k]
[X_{ij},X_{ik}] \right),  \\
&X_{ij} = S_{ij}T_\pi(r_{ij})+\oss Y_\pi(r_{ij}),
\eeqa
with strength $A_{2\pi}$. The functions $T_{\pi}(r_{ij})$ and $Y_{\pi}(r_{ij})$
describe the radial shapes of the one-pion exchange tensor and Yukawa
potentials. These functions are calculated using the average value
of the pion mass and include the short-range cutoffs used in the
Argonne $v_{18}$ NN interaction. The term denoted by $V^R_{ijk}$ is purely
phenomenological, and has the form:
\begin{equation}
V^{R}_{ijk}=U_0\sum_{cyc}T_\pi^2(r_{ij})T_\pi^2(r_{ik}).
\end{equation}
This term is meant to represent the modification of N$\Delta$- and 
$\Delta\Delta$-contributions in the two-body interaction 
by other particles in the medium, and also 
accounts for relativistic effects. The spin-isospin 
dependence of these effects is neglected.

The two parameters $A_{2\pi}$ and $U_0$ are chosen to yield the observed
energy of $^3$H and the equilibrium density of nuclear matter,
$\rho_0=0.16$ fm$^{-3}$. The parameters of Model VII, used by WFF, were
determined from approximate variational calculations using the older
Argonne $v_{14}$ NN interaction, and have the values:
$A_{2\pi}=-0.0333$ MeV and $U_0=0.0038$ MeV. The parameters of Model IX,
$A_{2\pi}=-0.0293$ MeV and $U_0=0.0048$ MeV, have been determined from
exact GFMC calculations of $^3$H and the
present approximate variational calculations of nuclear matter with
the Argonne $v_{18}$ NN interaction. Most of the difference between
the parameters of Models VII and IX is due to the use of exact,
rather than approximate calculations of $^3$H in the latter.
The parameters of Model VIII, also determined via exact
$^3$H calculations, albeit with the
older Argonne $v_{14}$ interaction, have the values: $A_{2\pi}=-0.028$ MeV 
and $U_0=0.005$ MeV, which are not very different from the present
Model IX parameters. The insensitivity of these parameters to the details
of the Urbana-Argonne models of $v_{ij}$ had been previously noted 
\cite{wiringa}.
For example, calculations of $^3$H and nuclear matter saturation density
using Model VII yielded similar results with the Urbana $v_{14}$ and
Argonne $v_{14}$ interactions, despite significant differences in the
tensor components of these interactions. We also note that the values
of $A_{2\pi}$ obtained from all these fits are close to $\sim-$0.03 MeV,
the strength predicted by the Fujita-Miyazawa model. More recent 
theoretical models of $V^{2\pi}_{ijk}$, such as the Tucson-Melbourne 
\cite{tuscon},
predict significantly larger strengths, $A_{2\pi}\sim-$0.063 MeV.
These models of $V^{2\pi}_{ijk}$ also have additional terms which are
neglected here.

The isotensor and isovector parts of $v_{ij}$, and the isovector part
of the kinetic energy K, are very weak and may presumably be treated
as first order perturbations. In first order, these terms do not 
contribute to the
energy of SNM, which has total isospin T=0.
Therefore, the SNM calculations, which neglect $v_{em}$ by definition,
are carried out with the isoscalar part of the Hamiltonian:
\begin{equation}
H_{SNM}=\sum_i -\frac{\hbar^2}{4} \left( \frac{1}{m_p} + \frac{1}{m_n} 
\right)\nabla_i^2
+\sum_{i<j}\sum_{p=1,14}v^p(r_{ij})O^p_{ij}
+\sum_{i<j<k} V_{ijk}.
\end{equation}
The isotensor and isovector terms do contribute to the energy of 
PNM. The isospin operators $\ott$,
$3\tau_{zi}\tau_{zj}-\ott$ and $\tau_{zi}+\tau_{zj}$
reduce to constants, respectively equal to 1,2 and $-2$, in PNM.
Hence, the full Hamiltonian, excluding $v_{em}$, has a simple
form for PNM, given by:
\begin{equation}
H_{PNM}=\sum_i -\frac{\hbar^2}{2m_n}\nabla_i^2
+\sum_{i<j}\sum_{q=1,7}\tilde{v}^{2q-1}(r_{ij})O^{2q-1}_{ij}
+\sum_{i<j<k} V_{ijk},
\end{equation}
with
\beqa
\tilde{v}^1 & = & v^1+v^2+2v^{15}-2v^{18} 
\label{pnmpot1}                           \\
\tilde{v}^3 & = & v^3+v^4+2v^{16}
\label{pnmpot2}                           \\
\tilde{v}^5 & = & v^5+v^6+2v^{17} 
\label{pnmpot3}                           \\
\tilde{v}^{2q-1} & = & v^{2q-1}+v^{2q} \ \mbox{for} \ q\ge 4 
\label{pnmpot4}
\eeqa
We note that $\left[ \ott,\boldtau_i\cdot\boldtau_k\right]=0$ in
PNM, thus the commutator term of $V^{2\pi}_{ijk}$ can be omitted
from H$_{PNM}$.

\section{VARIATIONAL CALCULATIONS}
The variational method developed in References \cite{pw,lp3,wff,lp} provides
a means to calculate the energy and wavefunction of the ground state 
of nuclear matter from realistic models of nuclear forces.
We outline the method here and describe additional developments in
Appendices A and B.

The variational wavefunction has the form:
\begin{equation}
\Psi_v = (S \prod_{i<j} F_{ij}) \Phi ,
\end{equation}
consisting of a symmetrized product of pair correlation operators
$F_{ij}$ operating on the Fermi gas wavefunction $\Phi$. In SNM,
the $F_{ij}$ include eight terms:
\begin{equation}
F_{ij}=\sum_{p=1,8} f^p(r_{ij})O^p_{ij} ,
\end{equation}
representing central, $\oss$, 
S$_{ij}$ and $\ols$
correlations with and without $\ott$ factors. In PNM, 
the $F_{ij}$ reduce to a sum of four terms with only odd $p\leq$7.

This wavefunction is clearly too simple to accurately describe the
ground state of nuclear matter. Monte Carlo studies of few-body nuclei 
\cite{apw} and $^{16}$O \cite{pwp} use additional three-body
correlation operators in the variational wavefunction. These additions
to the wavefunction lower the ground state energy of $^{16}$O by
$\sim$1 MeV/nucleon. Attempts to include three-body correlation
operators in variational calculations of SNM are currently in 
progress \cite{wu}.
Exact GFMC calculations have now been carried
out for nuclei containing up to seven nucleons \cite{pud1,pud3}. The 
ground state energy of $^7$Li obtained with variational wavefunctions
including three-body correlation operators is greater than the exact
value by $\sim$0.7 MeV/nucleon. From these results, we estimate that
the present $\Psi_v$ may underbind SNM
by a few MeV. In contrast, the three-body correlations have a
much smaller effect on the energy of pure neutron drops \cite{pscppr}.
The variational energy of a drop with eight neutrons, calculated with
the simple $\Psi_v$, is greater than the
exact value by $\sim$0.5 MeV/nucleon. Thus, we believe the energies
calculated in the present work to be relatively more accurate for
PNM than for SNM. This result is to be expected, since SNM has strong
tensor correlations in two-body np states, with isospin T=0, and in
three-body nnp and pnn clusters with T=$\frac{1}{2}$.

Despite the aforementioned shortcomings, the simple $\Psi_v$ having
only pair correlation operators describes
the gross features of the nuclear wavefunction rather well.
For example, the spin-isospin dependent two-nucleon distribution 
functions calculated in
this approximation are close to the exact distribution functions
\cite{pud2}. 
We note that the magnitude of the ground state energy is generally
much smaller than those of the positive kinetic and negative interaction
energies. The magnitude of the error in the ground state energy is
typically only a few percent of the interaction energy.
We can therefore study the main features of the spin-isospin structure of
PNM and SNM using the simple $\Psi_v$ described above.

The correlation operators $F_{ij}$ are determined from Euler-Lagrange 
equations \cite{lp3} that minimize the two-body cluster contribution
of an interaction $(\bar{v}-\lambda)$, where:
\beqa
\bar{v}_{ij} & = & \sum_{p=1,14}\alpha^p v^p(r_{ij}) O^p_{ij}, \\
\lambda_{ij} & = & \sum_{p=1,8}\lambda^p(r_{ij})O^p_{ij}.
\eeqa
The variational parameters $\alpha^p$ are meant to simulate the
quenching of the spin-isospin interaction between particles i and j,
due to flipping of the spin and/or isospin of particle i or j via
interaction with other particles in matter \cite{wp}. We use:
\beqa
\alpha^p & = & 1 \ \mbox{for} \  p=1 \ \mbox{and} \ 9  \\
\alpha^p & = & \alpha \ \mbox{otherwise},
\eeqa
since the operators 1 and $L^2$ ($p=1$ and 9) are independent of spin-isospin.
The $\alpha^p$ of the $\ols^2$ interactions ($p=13,14$) are
also set to unity. The $\ols^2$ operator has a significant $L^2$ part,
which should not be quenched, and quenching the remainder of
$v^{p=13}$, along with $v^{p=14}$, does
not lower the variational energy.
The $\lambda^p(r)$ simulate screening effects in matter
and are determined by the healing distances $d^p$ of the correlation
functions $f^p$:
\begin{equation}
f^p(r>d^p) = \delta_{p1}.
\end{equation}
The radial dependence of $\lambda^p(r>d^p)$ is determined from the
above constraint. The $\lambda^p(r<d^p)$ are constants chosen
to make the gradients of $f^p$ at $d^p$ equal to zero. The energy of
nuclear matter is particularly sensitive to the range of the tensor
and central correlations. For simplicity we assume:
\beqa
d^p & = & d_t \ \mbox{for} \ p=5,6 \\
d^p & = & d_c \ \mbox{for} \ p \neq 5,6.
\eeqa
The $F_{ij}$, and consequently $\Psi_v$, thus depend on 
three variational parameters:
$\alpha$, d$_c$ and d$_t$, the values of which are determined by
minimizing the energy. Two additional parameters used by
WFF have only a small influence on the energy.

It is convenient to divide the interaction and correlation operators
as follows:
\beqa
&v_{ij}= v_{s,ij} + v_{b,ij} + v_{q,ij} , \\
&F_{ij} = F_{s,ij} + F_{b,ij} .
\eeqa
The static parts $v_{s}$ and $F_{s}$ involve terms with 
momentum independent operators $O^{p=1-6}_{ij}$. The spin-orbit terms,
with $p=$7 and 8 are included in $v_b$ and $F_b$, while terms quadratic
in $L$, with $p=$9 to 14 are included in $v_q$. The two-body (2B) cluster
contributions involve expectation values of the operators:
\beqa
f^q(r_{ij}) O^q_{ij} v^p_{ij} O^p_{ij} f^{q'}(r_{ij}) O^{q'}_{ij}, \\
f^q(r_{ij}) O^q_{ij} \nabla^2_{ij} f^{q'}(r_{ij}) O^{q'}_{ij},
\eeqa
in plane-wave states. We divide these contributions into five
parts: the static parts, $\langle v \rangle$-2B-s 
($\langle T \rangle$-2B-s) include interaction
(kinetic) energy contributions for $q,p$ and $q'$ ($q$ and $q'$) $\leq$ 6;
the spin-orbit parts, $\langle v \rangle$-2B-b and 
$\langle T \rangle$-2B-b include terms with $p\leq$ 8,
and one or more of the $p,q,$ and $q'$ equal to 7 or 8; and the
quadratic part, $\langle v \rangle$-2B-q includes all terms with $p\geq$ 9.

The results obtained with various Hamiltonians for SNM at 
k$_f$=1.33 fm$^{-3}$ and 1.6 fm$^{-3}$ are listed in table~\ref{tabenm}. 
These Hamiltonians include the kinetic energy plus Urbana $v_{14}$ $v_{ij}$
(U14), Argonne $v_{18}$ $v_{ij}$ (A18) and Argonne $v_{18}$ $v_{ij}$
plus Model IX $V_{ijk}$ (A18+IX). The Fermi kinetic energy is 
listed as $\langle T \rangle$-1B, the one-body cluster 
contribution to the kinetic energy.

Variational cluster Monte Carlo studies of $^8$n-drops \cite{pscppr}
and $^{16}$O \cite{pwp,ppc}, including up to five-body cluster
contributions to the ground state energy, indicate that the convergence
of the cluster expansion is not particularly good in either case.
For example, the one- through five-body clusters 
contribute 12.9, $-$54.5, 11.1, $-$3.8, and 1.1 MeV respectively
to the ground state
energy of the $^8$n-drop. This result does not contradict the
earlier lowest-order constrained variational calculations of
neutron gas \cite{sp,p}, which retained only the $\langle T \rangle$-1B 
and all 2B contributions. The aim of the earlier studies was to
obtain estimates of the ground state energy; the $d_c$ and $d_t$
were not determined variationally, but rather were fixed by 
constraints. In the present work, $\Psi_v$ is determined variationally
to study the structure of SNM and PNM.
The larger optimum values of $d_c$ and $d_t$
lead to the significant size of many-body (MB) clusters of
three or more particles.

According to Table~\ref{tabenm}, the kinetic energy and static interactions
and correlations make large contributions to the 1B and 2B
cluster energies. We expect MB cluster contributions containing
$\nabla_i^2$ or $v_{s,ij}$ and $F_s$ correlations to be important.
In this work, as in earlier studies by FP and WFF,
these contributions are calculated using chain summation
methods \cite{pw}. 
The following three improvements were made by WFF in the basic VCS
calculation used by FP:

(i) The kinetic energy can be calculated using different expressions
related by integration by parts. If all MB contributions are calculated,
these expressions yield the same result. However, they yield different
results when only selected parts of the MB clusters are summed by VCS
techniques. Studies of atomic helium liquids with
VCS methods \cite{um} find the exact result to be between
the energies obtained using the Jackson-Feenberg (JF) and 
Pandharipande-Bethe (PB) expressions. Both the JF
and PB energies were calculated by WFF, who took the average of the two as 
the result, and half the difference as an estimate of the error.

(ii) The pair distribution function $g^c(r)$ is defined such that
$\rho g^c(r)$ represents the probability of finding a particle
at a distance r from a given particle. Conservation of number of
particles then implies:
\begin{equation}
I_c = \rho \int d^3r (1-g^c(r)) = 1,
\end{equation}
for both SNM and PNM. Also, since SNM has total isospin T=0,
we have the following identity for expectation values of
$\ott$ operators:
\begin{equation}
I_{\tau} = \frac{1}{A} \langle 0| \sum_{i,j=1,A} \ott |0\rangle 
= -3.
\end{equation}
The values of I$_c$ and I$_\tau$ calculated using the VCS
method are generally within a few percent of these exact results.
However at small densities, the correlations become large due to
the bound deuteron and virtual bound $^1$S$_0$ states, causing
I$_c$ and I$_\tau$ to deviate from their exact values by more
than 10\% in some regions of the $d_c$, $d_t$, $\alpha$ parameter space.
Deviations of this size can also occur at large densities.
Excursions into such regions of the parameter space are
curtailed in SNM by minimizing:
\begin{equation}
\left< H \right> + \Lambda \left[ (I_c-1)^2 +(\frac{1}{3}I_\tau+1)^2
\right]
\end{equation}
with $\Lambda$ chosen to keep $I_c$ and $I_\tau$ within 10\% of
their exact values. Only the $\Lambda(I_c-1)^2$ constraint is
applicable for PNM.

(iii) In their calculation, WFF
added the leading multiple-operator chain contributions
to those summed via VCS. At $\rho_0$, these terms contribute less
than 1 MeV/nucleon, while at $4\rho_0$, they may contribute a few
MeV/nucleon. Including these terms, WFF estimate the error in
the calculation of the MB contributions due to kinetic energy,
$v_s$ and $F_s$ to be $\sim$ 0.5 MeV/nucleon at $\rho_0$, and
$\sim$ 5 MeV/nucleon at $6\rho_0$ in SNM; the corresponding 
error in PNM is lower still. We note that these errors
are negligible compared to the error implicit in the use of the
simple $\Psi_v$ given by Eq.~(3.1).

The computer program used to carry out the present calculations
is built upon that of WFF and retains all the improvements
made by them.

The contribution of MB clusters involving spin-orbit correlations
and interactions is calculated using methods developed by
Lagaris \cite{lp}. Separable three-body clusters, with correlations
between pairs ij and ik, but not between jk, make the dominant
contribution to $\langle v+T \rangle$-MB-b, via $v_{ij}$ or $\nabla_i^2$
terms in the Hamiltonian. Chain diagrams with correlations
between all three pairs ij,ik and jk were found to make a smaller
contribution. We therefore sum all three-body separable contributions
to $\langle v+T \rangle$-MB-b, and estimate only the leading 
chain contributions.

Lagaris also calculated part of the three-body separable contribution
to $\langle v \rangle$-MB-q using the U14 Hamiltonian \cite{lp3}. The $L^2$ and 
${\bf (L\cdot S)}^2$ interactions that contribute to 
$\langle v \rangle$-MB-q are significantly
stronger in the A18 model than in the U14 model. We
therefore include a more complete calculation of three-body 
separable terms and leading central chain contributions to
$\langle v \rangle$-MB-q in the present work.
These calculations are outlined in appendix~A. 
The present calculation of $\langle v \rangle$-MB-q and the earlier
calculation by Lagaris yield similar results for the U14 Hamiltonian:
1.61 (3.91) MeV versus 1.35 (3.0) MeV at $k_F=$ 1.33 (1.6) fm$^{-3}$.
The difference between the the two calculations is significant
for the the A18+IX Hamiltonian, which makes much greater contributions 
to $\langle v \rangle$-MB-q (Table~\ref{tabenm}).

We have summed the contributions of leading three-body and more
than three-body diagrams to the expectation values of
the $V^{2\pi}$ and $V^R$ static three-body interactions, using 
VCS methods described in Reference \cite{cpw}.
The A18 interaction induces stronger correlations than U14, as is
evident by comparing T-2B-s for the two interactions in Table~\ref{tabenm}.
The larger correlations lead to  comparatively larger 
$\left<V^{2\pi}\right>$ and $\left<V^R\right>$.
The $\left<V^{2\pi}\right>$ in U14+VII \cite{spw} and A18+IX models
equal $-$3.49 and $-$3.60 MeV respectively at $k_F$=1.33 fm$^{-3}$,
despite the smaller strength of $V^{2\pi}$ in Model IX
($A_{2\pi}= -0.0293$ MeV) compared to Model VII ($A_{2\pi}=-0.0333$ MeV).
The $\left<V^R\right>$ is larger for  A18+IX (6.33 MeV at
$k_F$=1.33 fm$^{-3}$) compared to U14+VII (3.99 MeV at $k_F$=1.33 
fm$^{-3}$). Only about half the increase in $\left<V^R\right>$ is
due to the larger strength of $V^R$ in Model IX ($U_0$=0.0048 MeV)
compared to Model VII ($U_0$=0.0038 MeV).

We note that the two-body correlations in this variational calculation
do not have the optimal form. The $F_{ij}$ are obtained by minimizing
the sum of 2B contributions of the potential $\bar{v}$ (Eq.~3.3),
with healing constraints imposed at $d_c$ and $d_t$. More general
correlations can be generated by separately minimizing the two-body cluster
contribution to each partial wave, specified by $l,S,J$ and the relative 
momentum  $k$ \cite{pb}.
Thus, the correlations $f(l,S,J,k)$ depend on all the quantum
numbers, and yield a lower 2B energy than the $F_{ij}$ operators
defined in Eq.~(3.2) for $\bar{v}$ with the same $d_c$ and $d_t$.
The MB contributions cannot be easily
calculated with the general $f(l,S,J,k)$, however.

The $F_{ij}$ operators provide a good approximation to the two-body
correlations in matter. The small differences between optimum
$f(l,S,J,k)$ and $F_{ij}$ can be accounted for by inclusion of the
second order two-particle, two-hole contribution, $\Delta E_2$,
calculated in correlated basis perturbation theory \cite{ffp,ff}.
We estimate this contribution as described
in Appendix~B, approximating the $\Delta E_2$
by the difference $\delta E_{2B}$ between the 2B cluster energies
calculated using $f(l,S,J,k)$ and $F_{ij}$.
The values of $\alpha, 
d_c, d_t$ are determined by minimizing the energy calculated from
the $F_{ij}$, and $\delta E_{2B}$, calculated for these optimum
$\alpha, d_c, d_t$, is perturbatively added to the energy.
The calculated values of $\delta E_{2B}$ (Table~\ref{tabenm}) for the
U14 Hamiltonian are slightly smaller in magnitude than the
$\Delta E_2$ values reported in Reference \cite{ffp}, for the
same interaction. The $\delta E_{2B}$ correction is larger for the
A18 and A18+IX Hamiltonians, which predict stronger correlations
in matter. The $\Delta E_2$ is known to be relatively larger
for the A14 interaction \cite{ff}, which induces
stronger correlations than the U14. Our best estimate of the
variational energy, given by $\frac{1}{2}(E_{PB}+E_{JF})+\delta E_{2B}$,
is also listed in Table~\ref{tabenm}.

Over a decade ago, Day and Wiringa \cite{dw} calculated the
ground state energies of SNM by means of the Brueckner-Bethe
method. Contributions of up to  four hole-line diagrams were
included, in order to reduce the estimated error in the calculated
energy to $\sim\pm$0.18(1.3) MeV at $k_f=1.33(1.6) $ fm$^{-1}$.
The calculations demonstrated that the three realistic, two-body
potentials used in that work, Bonn (1975) \cite{bonn1,bonn2}, Paris 
\cite{paris} and A14, yielded essentially the same
energy of $\sim -14.9$ MeV at $k_f=1.33 $ fm$^{-1}$. 
At $k_f=1.6 $ fm$^{-1}$, the Paris and Argonne $v_{14}$ models
yielded similar energies of $\sim -17.8$MeV, while the Bonn
result was somewhat higher, at $\sim -16.8$MeV.
Our results with U14 and A18 (Table~\ref{tabenm}) are close to the earlier
results at $k_f=1.33 $ fm$^{-1}$, and are about 1 MeV lower at
$k_f=1.6 $ fm$^{-1}$. As with the results of Day and Wiringa, the
present results show a remarkable model independence of
the energy of SNM for $k_f\leq1.6 $ fm$^{-1}$, calculated from realistic
two-nucleon interaction models. The differences in the energies
obtained using the different models are smaller than the estimated 
error in the many-body calculations.

The energies of SNM and PNM calculated using the A18+IX Hamiltonian
appear in Fig.~\ref{esnm} and Fig.~\ref{epnm},
along with the results obtained by WFF using A14+VII. 
As discussed in detail below, there appears to be a phase transition
in both SNM and PNM with the present Hamiltonian. The curves
marked LDP and HDP show the energies obtained for the low and
high densitiy phases, respectively.

At larger densities, the A18+IX energies are
significantly higher than the A14+VII energies. A major part of 
this difference can be attributed to the difference between
$V_{ijk}$ in Models VII and IX. For example, the contribution of
Model IX exceeds that of Model VII by 8.2 (4.5) MeV to SNM (PNM)
energy at $\rho=0.32 $ fm$^{-3}$. The Model VII is unrealistic, however,
as it overbinds $^3$H and $^4$He.
The remainder of the difference between A18+IX and A14+VII is
due to the stronger momentum-dependent part, $v_{q,ij}$ of A18,
which makes significant contributions at larger densities.

In the early 1980s, FP added phenomenological density-dependent terms
to the U14 Hamiltonian, and adjusted their parameters to reproduce
the empirical equilibrium density, energy and compressibility of SNM.
These results also appear in Fig.~\ref{esnm} and Fig.~\ref{epnm}.
Energy density
functionals based on the FP results for E($\rho$) of SNM and PNM
reproduce the binding energies of nuclei from $^{16}$O to
$^{208}$Pb rather well \cite{ravenhall}.

The minimum energy calculated for SNM with the present variational
wavefunction and the A18+IX Hamiltonian is $-$12 MeV, compared to
the empirical value of $-$16 MeV. As previously mentioned, results
of variational and exact calculations of light nuclei 
suggest that including three-body correlations in the wavefunction
could lower the variational bound on the energy 
by more than 1 MeV, and that the true 
energy may be more than 1 MeV lower still. Thus, the underbinding
of SNM due solely to deficiencies in the A18+IX 
model of nuclear forces is probably much
less than 2 MeV. This underbinding is a very small fraction of
the total potential energy of SNM at $\rho_0=0.16 $ fm$^{-3}$,
which is $\sim-$50 MeV for this Hamiltonian.

At approximately twice the equilibrium density, we observe
what appears to be a transition in SNM due to a change in
$d_t$, the range of tensor correlations. The energy of SNM
is shown in Fig.~\ref{edtsnm} as a function of density and of $d_t/r_0$,
where $r_0$ is the unit radius defined by
\begin{equation}
\frac{4\pi}{3}r_0^3 \rho=1.
\end{equation}
The energies in Fig.~\ref{edtsnm} have been minimized with respect to
variations in the other two parameters, $\alpha$ and $d_c$, at
each $\rho$ and $d_t/r_0$. For $\rho<0.32 $ fm$^{-3}$, the minimum
occurs at $d_t/r_0 \sim 4$, whereas for $\rho>0.32 $ fm$^{-3}$, it
shifts to $d_t/r_0 \sim 6$. The upper line in Fig.~\ref{esnm},
labeled LDP represents the minimum energies for $d_t/r_0 \sim 4$, while
the lower line labeled HDP gives the energies of the minima at
larger values of $d_t/r_0$.

This transition is probably related to pion condensation
\cite{wff,migdal,sp}, and its spin-isospin structure is
discussed in the following sections. The transition does not occur
in similar calculations of SNM using either the U14 or
A14 and is thus sensitive to the two-nucleon interaction model.
The Fujita-Miyazawa two-pion exchange three-nucleon 
interaction is a necessary ingredient for the transition
to occur in SNM.
A similar transition occurs for PNM (Fig.~\ref{epnm}) at a lower
density ($\sim 0.2 $ fm$^{-3}$) with both the A14+VII and
A18+IX Hamiltonians, but it is not observed for U14+VII.
The $E(\rho,d_t/r_0)$ for PNM with A18+UIX appear in Fig.~\ref{edtpnm}.
Unlike the A14 and U14 interactions,
the transition persists in PNM with A18, in the absence
of the Fujita-Miyazawa interaction,
although it occurs at a much higher density ($\sim 3.5 \rho_0$)
in that case.

The contributions to $E(\rho,d_t/r_0)$ of SNM at selected
values of $\rho$ and $d_t/r_0$ are listed in Table~\ref{tabedt} for
the A18+IX model. The interactions $v^{t\tau}$,
$v^{\sigma\tau}$ and $v^c$, associated with the operators
$S_{ij}\ott$, $\oss\ott$
and 1, make the largest contributions to the energy of SNM.
These contributions are listed separately, along with the
contribution of the entire $v$. The $v^{t\tau}$
and $v^{\sigma\tau}$ contributions come mainly from the one-pion
exchange interaction. The expectation values of $v^{\sigma\tau}$,
$V^{2\pi}$ and many-body cluster contributions are significantly
different in the LDP and HDP.
The pion-exchange contributions to the energies of PNM and SNM
are listed in Table~\ref{vpinpi} for the LDP and HDP. 
The OPEP used to calculate $\langle v^\pi \rangle $ includes a 
$\pi N N$ dipole form factor
with a 5 fm$^{-1}$ cutoff as described in Section VI. The $V^{2\pi}$
is that given by Model IX.
The results indicate that for SNM the change in the pion-exchange
contribution between the two phases comes overwhelmingly from
the $V^{2\pi}$, which is required to produce the
phase transition. However, for PNM a significant part
of the enhancement in the pion-exchange contribution comes
from $v^\pi$, thus indicating the diminished importance
of the three-body interaction to the phase transition, and the
persistence of the transition in the absence of $V^{2\pi}$.

In the present calculation, the transition in SNM is of
first order, as is evident in Fig.~\ref{esnm}. Thus, it is difficult to
obtain information about the HDP by studying the LDP. The order
of the corresponding transition in PNM is not as evident in
Fig.~\ref{epnm}, though it appears to be of first order as well.

\section{PAIR DISTRIBUTION FUNCTIONS}
The two-body density $\rho_2^p(r)$, associated with the operator
O$^p_{ij}$, is defined \cite{pwp} such that:
\begin{equation}
\langle 0| \sum_{i\neq j =1,A} B(r_{ij}) O^p_{ij} |0\rangle =
A\int d^3r B(r) \rho_2^p(r),
\end{equation}
for any function B(r$_{ij}$). This relationship is used to calculate the
expectation values of the pair interactions $v^p(r_{ij})O^p_{ij}$
in the ground states of nuclei and nuclear matter, denoted by
$|0\rangle$. The $\rho_2^p$ associated with the static operators
1, $\ott$, $\oss$, $\oss\ott$, $\os$ and $\os\ott$, are 
denoted by $\rho_2^c$, $\rho_2^\tau$,
$\rho_2^\sigma$, $\rho_2^{\sigma\tau}$, $\rho_2^t$ and
$\rho_2^{t\tau}$. All properties of SNM
discussed in this and the following two sections are obtained from
these six two-body densities. In the case of PNM, we disregard
isospin and consider only $\rho_2^c$, $\rho_2^\sigma$ and $\rho_2^t$.
The central two-body density, $\rho_2^c$, is proportional to the
probability of finding a pair of particles separated by a distance r,
and  asymptotically approaches the matter density, $\rho$, as 
r$\rightarrow \infty$. All other $\rho_2^p$ are associated with 
spin-isospin correlations and must therefore
vanish as r$\rightarrow \infty$.

The static $\rho^p_2$, calculated  for SNM at the equilibrium
density of $\rho=$ 0.16 fm$^{-3}$, appear in Fig.~\ref{snmrho16}.
The large magnitudes of the $\rho_2^{p>1}$ indicate
that short-range correlations
in SNM are strongly spin-isospin dependent. The nature of these
correlations is more evident in the pair distribution functions
$\rho^{(2)}_{T,S,M}(\bf r)$, which are proportional to the probability
of finding a pair of nucleons with total isospin T, spin S and
spin projection M, as a function of {\bf r}. Forest {\it et al.}
recently studied these densities in light nuclei \cite{forest}.

The $\rho^{(2)}_{T,0,0}(r)$, in S=0,M=0 states, are spherically
symmetric and are obtained from the $\rho_2^p$'s using spin-isospin
projection operators, such as $P_{S=0} = \frac{1}{4}(1-\oss)$.
In SNM,
\beqa
\rho^{(2)}_{1,0,0}(r) & = & \frac{1}{16}\left[3\rho^c_2(r)+\rho^\tau_2(r)
-3\rho^\sigma_2(r)-\rho^{\sigma\tau}_2(r)\right], \\
\rho^{(2)}_{0,0,0}(r) & = & \frac{1}{16}\left[\rho^c_2(r)-\rho^\tau_2(r)
-\rho^\sigma_2(r)+\rho^{\sigma\tau}_2(r)\right]. 
\eeqa
Whereas in PNM,
\beq
\rho^{(2)}_{0,0}(r)=\frac{1}{4}\left[\rho^c_2(r)-\rho^\sigma_2(r)\right].
\eeq

The pair distribution functions in S=1 states have a quadropolar
deformation due to the tensor interaction. The $\rho^{(2)}_{S,M}$
in PNM are given by:
\beqa
\rho^{(2)}_{1,0}({\bf r}) & = & C_0(r) -2C_2(r)P_2(cos\theta), 
\label{rtsm0}\\
\rho^{(2)}_{1,\pm 1}({\bf r}) & = & C_0(r) +C_2(r)P_2(cos\theta).
\label{rtsm1}
\eeqa
The expectation value of $B_1(r_{ij})P_{S=1}$, where 
$P_{S=1} = \frac{1}{4}(3+\oss)$,
can be calculated from either the two-body densities $\rho_2^{(p)}$
or from the distributions $\rho^{(2)}_{1,M}$. Equating the two
results, we find:
\beq
C_0(r)= \frac{1}{3} \times \frac{1}{4}
\left[3\rho^c_2(r)+\rho^\sigma_2(r)\right].
\eeq
A similar calculation of the expectation value of $B_t\os$ yields:
\beq
C_2(r)=\frac{1}{12}\rho^t_2(r).
\eeq

In the case of SNM, the $C_0(r)$ and $C_2(r)$ in T=0,1 states 
are obtained from the expectation
values of $B_1(r_{ij})P_{S=1}$ and $B_t\os$, multiplied by
isospin-projection operators, $P_{T=0,1}$. For the T=0,S=1
distribution functions, we obtain:
\beqa
C_0(r) & = & \frac{1}{3} \times \frac{1}{16}
\left[3\rho^c_2(r)-3\rho^\tau_2(r)+\rho^\sigma_2(r)-\rho^{\sigma\tau}_2(r)\right], \\
C_2(r) & = & \frac{1}{12} \times \frac{1}{4} 
\left[\rho^t_2(r)-\rho^{t\tau}_2(r)\right]. 
\eeqa
The corresponding T=1 expressions are given by:
\beqa
C_0(r) & = & \frac{1}{3} \times \frac{1}{16}
\left[9\rho^c_2(r)+3\rho^\tau_2(r)+3\rho^\sigma_2(r)+\rho^{\sigma\tau}_2(r)\right], \\
C_2(r) & = & \frac{1}{3} \times \frac{1}{16} 
\left[3\rho^t_2(r)+\rho^{t\tau}_2(r)\right].
\eeqa

The extrema of $\rho^{(2)}_{T,1,M}(r,\theta)$ at fixed $r$ occur for $M=0$ at
$\theta=0$ and $\theta=\frac{\pi}{2}$. These extrema are plotted as
a function of $r$ in Figs. \ref{snmrts01},\ref{snmrho36} and \ref{pnm111020}.
From Eqs.~\ref{rtsm0}-\ref{rtsm1}, it follows that:
\beqa
\rho^{(2)}_{T,1,\pm 1}(r,\theta =0) &=& \rho^{(2)}_{T,1,0}(r,\theta =\pi/2),\\
\rho^{(2)}_{T,1,\pm 1}(r,\theta =\pi/2) &=&
\frac{1}{2}\left(\rho^{(2)}_{T,1,0}(r,\theta =0) +\rho^{(2)}_{T,1,0}(r,\theta 
=\pi/2) \right).
\eeqa
We also note that in SNM:
\beq
\rho^{(2)}_{T,S,M}(r\rightarrow\infty)=\frac{2T+1}{16}\rho \ .
\eeq

The $\rho^{(2)}_{0,1,0}(r,\theta=0,\pi/2)$ are shown in Fig.~\ref{snmrts01}
for SNM at $\rho=0.16$ fm$^{-3}$, along with the corresponding
functions for $^2$H, $^4$He and $^{16}$O, from Ref.~\cite{forest}.
At small $r$, this density is large for $\theta=\pi/2$, where
the OPE tensor potential is attractive, and small for $\theta=0$,
where the OPEP is repulsive. Thus, according to Ref.~\cite{forest},
equidensity surfaces having $\rho^{(2)}_{0,1,0}> 0.01$ fm$^{-3}$
are toroidal in shape. The ratio $\rho^{(2)}_{0,1,0}(r,\theta=\pi/2)
/\rho^{(2)}_{0,1,0}(r,\theta=0)$ is a measure of the strength of
tensor correlations in T=0 states.
For the maximum possible tensor correlations, 
$\rho^{(2)}_{0,1,0}(r,\theta =0)$ is negligible compared to
$\rho^{(2)}_{0,1,0}(r,\theta =\pi/2$). Fig.~\ref{snmrts01} thus
indicates that the tensor correlations in T=0 states,
in nuclei and in nuclear matter, have near the maximum possible
strength at $r\sim 1$ fm.
The peak value of $\rho^{(2)}_{0,1,M}({\bf r})$
is almost 2.5 times the asymptotic value of 0.01 fm$^{-3}$ at
$\rho=0.16$ fm$^{-3}$. The spherically symmetric two-body densities
for SNM and nuclei in the T=1, S=0 channel ($\rho^{(2)}_{1,0,0}({\bf r})$)
are shown in Fig.~\ref{snmrts10}.
These distributions peak at $r\sim 1$, where the
nuclear force is most attractive, and the peak value is about 1.5
times the asymptotic value. Both $\rho^{(2)}_{1,0,0}({\bf r})$
and $\rho^{(2)}_{0,1,M}({\bf r})$ are supressed near $r\sim 0$
by the repulsive core in the NN interaction.

According to Ref.~\cite{forest}, the $\rho^{(2)}_{0,1,M}({\bf r})$
and $\rho^{(2)}_{1,0,0}({\bf r})$ have universal shapes in light
nuclei at small $r$.
In Figs.~\ref{snmrts01} and \ref{snmrts10}, we have scaled the
densities in light nuclei such that their maximum values equal
those of SNM.
The two-body density distributions in SNM appear to have
nearly the same shape as those in light nuclei for $r\alt 1.5$ fm.
However, significant differences occur for $r\agt 2$ fm. Note that the
average interparticle distance in SNM at $\rho=0.16$ fm$^{-3}$ is
also $\sim 2$ fm.

Forest {\it et al.} argue in Ref.~\cite{forest}, that the ratio, $R_{Ad}$,
of the maximum values of $\rho^{(2)}_{0,1,M}$ in a nucleus A
and the deuteron provides a good approximation to the
Bethe-Levinger factor, $L_{A}$, of the nucleus A. The pion
and photon absorbtion cross sections in light nuclei scale with
$R_{Ad}$. The calculated value of $R_{Ad}$ for SNM at equilibrium
density is 1.59A, which corresponds to $L_A=$6.36. 

We observe interesting changes in the two-body densities between
the low- and high-density phases in SNM and PNM.
The existence of a pion condensate is indicated in the
HDP, as discussed in the following sections. The
$\rho^{(2)}_{T,S,M}({\bf r})$ in SNM at $\rho=0.36$ fm$^{-3}$
in the LDP, with $d_t/r_0=4$, and in the HDP, with
$d_t/r_0=6$ are shown in Fig.~\ref{snmrho36}.
The analogous $\rho^{(2)}_{S,M}({\bf r})$
in PNM at $\rho=0.20$fm$^{-3}$ are shown in Fig.~\ref{pnm111020}.
The differences between pair densities in LDP and HDP are more
pronounced in PNM. In both PNM and SNM, we find that the
long-range part of the tensor correlations is enhanced. In PNM,
the S=0 $\rho^{(2)}_{0,0}$ is suppressed in the HDP, whereas in SNM,
$\rho^{(2)}_{1,0,0}$ and $\rho^{(2)}_{0,1,M}$ are suppressed
and $\rho^{(2)}_{0,0,0}$ is enhanced.

In Migdal's approach \cite{migdal}, the transition to the pion
condensed phase is inhibited by a positive, short-range
$\oss\ott$ two-nucleon interaction $v^{\sigma\tau}(r)$, represented
by the Landau parameter $g^\prime$.
The eigenvalues of $\oss\ott$ are -3, -3, 1 and 9 in T,S=1,0; 0,1;
1,1; and 0,0 states, respectively. Therefore, a strong positive
$v^{\sigma\tau}(r)$ favors the LDP, which has larger pair
densities in T,S=1,0 and 0,1 states, and a smaller pair density
in T,S=0,0. Similarly, in PNM, a large positive $\tilde{v}^\sigma$,
given by Eq.~\ref{pnmpot2}, favors the LDP. The $v^{\sigma\tau}(r)$
and $\tilde{v}^\sigma(r)$ in the U14, A14 and A18 models
are shown in Fig.~\ref{vst}. The positive $v^{\sigma\tau}$ and
$\tilde{v}^\sigma$ of U14 prevent a transition to the HDP in
both SNM and PNM. In A14, $\tilde{v}^\sigma$ becomes negative
at small $r$, and PNM is thus predicted by WFF to undergo a
transition, while in A18, both $v^{\sigma\tau}$ and $\tilde{v}^\sigma$
change sign and thus cannot prevent the transition in either SNM
or PNM. The Urbana-Argonne potentials have similar forms, but
are fit to different data sets. Whereas the U14 and A14 were fit
to n-p phase shifts available in the late 1970s and early 1980s, respectively,
the A18 is fit directly to the 1994 Nijmegen p-p and n-p
scattering database. Also, a much better fit was
achieved by A18 ($\chi^2$ per datum = 1.09). Thus, it is
likely that A18 provides a more accurate representation
of the NN interaction within the Urbana-Argonne framework.

The $v^{\sigma\tau}(r)$ predicted by $\pi-$ and $\rho-$exchange
potentials is positive for point-like nucleons. However, these
meson-exchange potentials also contain a negative $\delta-$function
term. It is possible that this term is broadened by the finite size
of nucleons, and that the total $v^{\sigma\tau}(r)$ changes sign at
small $r$. The one-meson exchange representation of the NN interaction
may not be reliable at small $r$, however.

The $v^{t\tau}\!(r)\os\ \ott$ interaction contains the main part of the
OPEP. At $r>1$ fm this interaction is essentially identical in the
U14, A14 and A18 models. At $r<1$ fm it is strongest in A14 and weakest
in U14.

\section{ISOVECTOR SPIN-LONGITUDINAL RESPONSE}

Migdal \cite{migdal} calculated the isovector spin-longitudinal
(IVSL) response  of nuclear matter using effective interactions.
The IVSL response is defined as:
\beqa
R_L(q,\omega) &=& \sum_I |\langle I|O_L({\bf q})|0\rangle|^2
\delta(\omega_I-\omega_0-\omega), \\
O_L({\bf q}) &=& \sum_{i=1,A} {\bf \boldsigma_i \cdot q \boldtau_i\cdot\hat{t}}
\ e^{i\bf{q\cdot r_i}},
\eeqa
where ${\bf \hat{t}}$ is a unit vector in isospin space, and 
$|0\rangle$ and $|I\rangle$ represent the ground and excited states
of the system, with energies $\omega_0$ and $\omega_I$, respectively.
SNM has zero isospin, and $R_L$ is therefore independent of
the direction of ${\bf \hat{t}}$.
The operator $O_L({\bf q})$ represents the coupling
of an external pion field to NM. In the case of PNM, we take
${\bf \hat{t}}$ to be in the z-direction, such that
$\boldtau \cdot {\bf \hat{t}}=-1$.
$O_L({\bf q})$ then represents the coupling
of a $\pi^0$ field to PNM. Migdal assumed that this response
would be dominated by a spin-isospin sound mode, and that the 
occurence of transition would be indicated by the vanishing of
the corresponding excitation energy.

Calculation of the response of NM from realistic interactions is
an extremely difficult task. However, it is well known \cite{pcpws}
that the sums and energy-weighted sums of response functions are
related to the two-body densities. For the IVSL response,
the sum and energy-weighted sum are defined as:
\beqa
Aq^2S_L(q) & = & \int_0^\infty R_L(q,\omega) d\omega, \\
Aq^2W_L(q) & = & \int_0^\infty R_L(q,\omega) \omega d\omega,
\eeqa
thus removing the dependence on the number of particles.
In SNM they are given by:
\beqa
S_L(q) & = & 1+\frac{1}{9}\int
\left[\rho^{\sigma\tau}_2(r)j_0(qr)-\rho^{t\tau}_2(r)j_2(qr)\right]d^3r, \\
W_L(q) & = & \frac{q^2}{2m} +\frac{1}{2}
\int\left[\sum_{p=2,6} D^p_L(r)\rho^p_2(r)\right]d^3r,
\label{wlq}
\eeqa
with the $D^p_L(r)$ tabulated in ref \cite{pcpws}. In the case
of PNM we obtain:
\beq
S_L(q)  =  1+\frac{1}{3}\int
\left[\rho^\sigma_2(r)j_0(qr)-\rho^t_2(r)j_2(qr)\right]d^3r,
\eeq
and
\beqa
D^\sigma_L(r) & = & -\frac{8}{3}\left[
\tilde{v}^\sigma(r)(1-j_0(qr))-\tilde{v}^t(r)j_2(qr)\right], \\
D^t_L(r) & = & -\frac{4}{3}\left[
\tilde{v}^t(r)(2+j_0(qr))-(\tilde{v}^\sigma(r)-2\tilde{v}^t(r))
j_2(qr)\right].
\eeqa
We note that terms with even $p$ do not occur in PNM, and the $\tilde{v}^p$
are defined as in Eq.~(\ref{pnmpot1}-\ref{pnmpot4}).
The energy-weighted sum of Eq.~(\ref{wlq}) contains only the 
contributions from the static parts of $v_{ij}$. These represent the
dominant contributions to the energy-weighted sums of responses to
electromagnetic fields \cite{sfp,schiav}, and presumably to $W_L$ as well.

The calculated values of $S_L(q)$ in SNM at $\rho=0.16$ fm$^{-3}$
and $\rho=0.36$ fm$^{-3}$, and in PNM at $\rho=0.16$ fm$^{-3}$ and
$\rho=$0.2 fm$^{-3}$, appear in  Fig.~\ref{slq}. At the higher densities,
results obtained for both the LDP and HDP are shown.
At equilibrium density, the $S_L(q)$ exhibits a small enhancement
in the $q\sim 2$ fm$^{-1}$ region. Indications of an enhancement
of the IVSL response have been observed in $(\vec{p},\vec{n})$
reactions \cite{tadd}. In the LDP, this enhancement grows slowly
with density. We predict a much larger enhancement 
at $q\sim 1.3$ fm$^{-1}$ in the HDP.

When the response is dominated by a single collective mode, i.e.
when only one of the states $|I\rangle$ largely contributes to the
sum in Eq.~(5.1), the energy of the collective state is given by
$W_L(q)/S_L(q)$. As an example, the energy of Feynman phonons
in atomic liquid $^4$He can be obtained from the $W_L/S_L$ ratio.
The spin-longitudinal response of nucleon matter probably has a large
spread in energy; nevertheless, we can define a mean energy of the
response as:
\beq
\bar{E}_L(q)=\frac{W_L(q)}{S_L(q)} \ .
\eeq
The resulting values appear in  Fig.~\ref{wos} for the cases
considered in Fig~\ref{slq}.

We note that in the LDP, $\bar{E}_L(q)$ is larger than
$q^2/2m$ in SNM as well as in PNM, indicating that the
nuclear interactions push the response to higher energies
on average.  
At $q\sim 0$ the IVSL response is almost entirely due to 
spin-isospin correlations, and $\bar{E}_L(q\sim 0)$ is
therefore large.
In Migdal's picture \cite{migdal}, the energy of the collective
spin-isospin sound wave with $q\sim 1.3$ fm$^{-3}$
decreases with increasing matter density,
and pion condensation occurs when the energy vanishes. In the
LDP, where $d_t\sim 4r_0$, the present calculations show that
$\bar{E}_L(q)$ does not decrease with density at any $q$. However,
the HDP has a lower $\bar{E}_L(q)$ than the LDP
in the vicinity of $q\sim 1.3$ fm$^{-1}$. It is likely that 
a part of the response at $q\sim 1.3$ fm$^{-1}$ shifts to
lower energies, or softens, as the system moves from the
LDP to the HDP.

\section{PIONIC INTERACTIONS AND EXCESS}
If the HDP is in fact a ``pion-condensed'' phase,
it must have associated with it an enhanced pion field 
and enhanced pion exchange interactions between the nucleons.
The interaction between two nucleons, due to the exchange of a
pion of momentum {\bf q}, is given by:
\beq
v^\pi_{ij}({\bf q}) = -\frac{f^2_{\pi NN}}{m^2_\pi(m^2_\pi+q^2)}
{\bf \boldsigma_i \cdot q \bf \boldsigma_j \cdot q}\ott
\ e^{i\bf{q}\cdot ({\bf r}_i-{\bf r}_j)}\Lambda^2(q),
\eeq
where $\Lambda(q)$ is the pion-nucleon form factor.
The expectation value of the interaction is trivially related
to the sum $S_L(q)$ of the IVSL response:
\beqa
\frac{1}{A}\left< \sum_{i<j}v^\pi_{ij}({\bf q})\right> & = &
-\frac{f^2_{\pi NN}}{m^2_\pi(m^2_\pi+q^2)}
\frac{3}{2} q^2(S_L(q)-1) \Lambda^2(q), \\
\frac{1}{A}\left< \sum_{i<j}v^\pi_{ij}({\bf q})\right> & = &
-\frac{f^2_{\pi NN}}{m^2_\pi(m^2_\pi+q^2)}
\frac{1}{2} q^2(S_L(q)-1) \Lambda^2(q),
\eeqa
in SNM and PNM, respectively.
The expectation value of the complete $v^\pi_{ij}$ is obtained 
by integrating these expressions over d{\bf q}:
\beqa
\frac{1}{A} \left< \sum_{i<j}v^\pi_{ij}\right> 
& = & \frac{1}{2A\pi^2}\int dq \ q^2 
\left< \sum_{i<j}v^\pi_{ij}({\bf q})\right> \nonumber \\
& = & \int dq \ \xi(q).
\label{vpiq}
\eeqa
The function $\xi(q)$:
\beq
\xi(q) = \frac{1}{2\pi^2}\frac{f^2_{\pi NN}}{m^2_\pi(m^2_\pi+q^2)}
q^4 \frac{3}{2} (1-S_L(q)) \Lambda^2(q)
\eeq
gives the pion-exchange nucleon-nucleon interaction (OPEP)
contribution as a function of q, the magnitude of the momentum
of the exchanged pions in SNM;
$\xi(q)$ for PNM is diminished by a factor of 3. The calculated values
of $\xi(q)$ for the cases previously discussed appear in 
Fig.~\ref{xiq} for SNM and PNM, where we have used:
\beq
\Lambda(q^2)=\frac{\lambda^2}{\lambda^2+q^2}
\eeq
with $\lambda$=5 fm$^{-1}$ as an illustration.
In the LDP, the attraction from the OPEP comes from a broad region
around $q\sim 3$ fm$^{-1}$, whereas in the HDP, the attraction is 
relatively more concentrated at $q\sim 1.5$ fm$^{-1}$.
Also, much of the repulsion due to low $q$ pions in the LDP is
absent in the HDP. The total OPEP contribution, obtained from
Eq.~(\ref{vpiq}), is listed in Table~\ref{vpinpi}.
In the case of SNM, the $\langle v^\pi \rangle$ is not very
different in the LDP and HDP; a much larger change occurs in
the $\langle V^{2\pi}\rangle$, also listed in table~\ref{vpinpi}.
In PNM, however, the difference in $\langle v^\pi \rangle$ is
more pronounced between the LDP and the HDP.

The difference in the expectation values of the pion number operator
in matter and for A isolated nucleons is called the pion excess
\cite{fpw}. The operator for excess pions of momentum {\bf q},
due exclusively to OPE interactions,
is given by:
\beq
\delta n^{(1)}_\pi(q)= -\frac{v^\pi(q)}{\surd m^2_\pi + q^2}.
\eeq
The distribution of excess pions is given by:
\beq
\eta(q)=-\frac{\xi(q)}{\surd m^2_\pi + q^2},
\eeq
which is shown in Fig.~\ref{etaq} for SNM and PNM. 
The $\eta(q)$ exhibits a sharp enhancement in the HDP 
at $q\sim 1.5$ fm$^{-1}$.

The integral of $\delta n^{(1)}_\pi(q)$, denoted by
$\left<\delta n^{(1)}_\pi\right>$,
is listed in Table~\ref{vpinpi}. 
The total $\left<\delta n_\pi\right>$ includes other contributions from
the NN$\rightarrow$N$\Delta$ and NN$\rightarrow \Delta\Delta$
transition interactions \cite{fpw}, which we have not calculated
here. These additional contributions take into account the
changes in the pion field due to N$\Delta$- and 
$\Delta\Delta$-box diagrams and due to the $V^{2\pi}_{ijk}$.

\section{CONCLUSIONS}
We have studied the short-range spin-isospin structure of SNM
and PNM, using one of the most accurate models of nuclear
forces currently available. For SNM at equilibrium density,
we predict a short-range structure very similar to that
found by Forest {\it et al.} \cite{forest} in light nuclei, at
interparticle distances $< 1.5$ fm. Symmetric Nuclear Matter
is bound by small 
localized regions of strong attraction in the NN potential
in T,S = 0,1 and 1,0 states. The two-nucleon densities are
found to have large overshoots, ranging up to 2.5 (1.5)
times the uncorrelated values (Figs.~\ref{snmrts01} and
\ref{snmrts10}) in the T,S = 0,1 (1,0) attractive regions.
The interaction, and consequently the pair density, in
T,S = 0,1 states is highly anisotropic due to the presence
of the OPE tensor force. The two-body density overshoots in
this state have femtometer-sized toroidal  structures
similar to those found in light nuclei; we therefore expect this feature
to occur in all nuclei. These short-range structures are not very
sensitive to the uncertainties in models of nuclear forces as
discussed in Reference~\cite{forest}.

We also find that the Argonne $v_{18}$ plus Urbana IX $V_{ijk}$
model of nuclear forces, which offers one of the best fits to
the Nijmegen NN scattering database, as well as to the
binding energies of light nuclei \cite{pud1}, predicts
that both SNM and PNM will undergo  transitions to  phases with
pion condensation at densities of $\sim 0.32$ fm$^{-3}$
and $\sim 0.2$ fm$^{-3}$, respectively. The occurrence of this
transition is sensitive to the short-range parts of the $\oss$ and
$\oss\ott$ NN interaction, as predicted by Migdal \cite{migdal}.
The transition does not occur with the older Urbana and Argonne 
$v_{14}$ models in SNM, while in PNM it occurs with the Argonne 
$v_{14}$, though not with the Urbana $v_{14}$.

It should be stressed that the present calculations of pion
condensation in the HDP are incomplete. Although we have used a variational
wavefunction of the same form to describe both the LDP and the HDP,
a different form permitting long-range order should be used for the latter.
A wavefunction describing a correlated liquid crystal, containing 
layers of spin-isospin ordered nucleons, may be more appropriate for
the HDP. Takatsuka {\it et al.} \cite{tttt} have used such a wavefunction,
denoted ASL for ``alternating spin layers''. A correlated spin-ordered solid
(SOS) wavefunction \cite{ps} has also been used in the past.
However, only lowest-order  variational \cite{ps} or G-matrix \cite{tttt}
calculations have been possible with these wavefunctions having
long-range order.
In view of the small difference between the energies of the LDP, and
the HDP, it may be advisable to use similar methods to calculate both.
The chain summation methods used here cannot be used with either ASL
or SOS wavefunctions. However, in future it may be possible to
calculate energies of the LDP as well as correlated ASL and SOS
phases, using cluster Monte Carlo methods \cite{pwp,ppc,pscppr}.

The indication of a phase transition obtained with the present calculation
should be reliable. The changes in the pair distributions, the IVSL
response and the pion fields strongly suggest that the HDP will
exhibit pion condensation upon inclusion of wavefunctions 
permitting long-range order. We hope that such a modification
will not significantly alter our results for the short-range structure
and energy of the HDP.
The momentum of the condensed pion field is $\sim 1.4$ fm$^{-1}$ at
the onset of the phase transition, which corresponds to a
wavelenth of 4.5 fm. The interlayer spacing, which equals half the
wavelength \cite{vrpsolvey}, is thus predicted to be $\sim 2.25$ fm
in both SNM and PNM. This spacing is larger than the spacing of
1.5 (1.7) fm between the layers of a simple cubic solid at
$\rho=0.32 (0.2)$ fm$^{-3}$.

Most calculations exploring the possibility of pion condensation in
matter explicitly consider the $\pi$N$\Delta$ coupling. The baryons in
ASL and SOS matter are taken to be superpositions of nucleon and
$\Delta$-states. The NN~$\rightarrow$~N~$\Delta$ and 
NN~$\rightarrow\Delta\Delta$ transitions are considered in these
approaches, which make the theory more difficult. In contrast, we
do not consider the $\Delta$-degree of freedom explicitly in the 
present work. The resulting effects, along with those due to
other mesons and nucleon resonances, are implicit in the two- and
three-nucleon potentials obtained by fitting experimental data.
Our Hamiltonian has only nucleon degrees of freedom, and its 
predictions can be calculated using a variety of many-body 
techniques. As previously mentioned, however, the predictions regarding
pion condensations are sensitive to the details of the
short-range NN interaction.

It is encouraging to note that our calculated density of
$\sim 0.32$ fm$^{-3}$ for the onset of pion condensation in SNM
is within the range 0.32 to 0.48 fm$^{-3}$ favored by calculations
using the ASL wavefunction \cite{tttt}. As with WFF,
we predict a lower transition density ($\sim 0.2$ fm$^{-3}$)
for PNM. This is due to the fact that in most realistic
models, $\tilde{v}^\sigma(r)$, the relevant interaction in PNM,
is softer than $v^{\sigma\tau}(r)$ (Fig.~\ref{vst}).
Calculations with ASL wavefunctions predict a higher density
of $\sim 0.5$ fm$^{-3}$ for the onset of neutral pion
condensation in PNM \cite{tttt}. The possibility of charged
$\pi^-$ and/or $K^-$ condensation in neutron star matter
at densities above our result of 0.2 fm$^{-3}$ for $\pi^0$ condensation
is currently being investigated by a number of researchers
\cite{pm,ky,kn}.

\acknowledgements
The authors thank J.L. Forest, S.C. Pieper and R.B. Wiringa for many
discussions and useful suggestions,
D.S. Lewart for help preparing the manuscript and figures, and 
C.M. Elliot for proofreading the manuscript.
This work was supported by the U.S.\ National Science
Foundation via grant PHY 94-21309. Calculations were performed on the
Cray C90 at the Pittsburgh Supercomputing Center.

\appendix
\section{THE TREATMENT OF $L^2$ AND ${\bf (L\cdot S)}^2$ INTERACTIONS}

The leading contribution to $\langle v \rangle$-MB-q 
is from the separable, direct three-body diagram shown in Fig.~\ref{diag}a. 
The earlier calculation of $\langle v \rangle$-MB-q 
by Lagaris \cite{lp3} with the
U14 potential included only the main term of this diagram,
namely the term with only central correlations between interacting
particles. While this approximation was justifiable for
U14, the much stronger $L^2$ and $({\bf L\cdot S})^2$ interactions of A18
(see Table~\ref{tabenm}) require calculation of additional terms. 
In the present work we include all relevant terms of the direct diagram
presented in Fig.~\ref{diag}a, and in addition consider:
i) the interacting exchange three-body separable diagram (Fig.~\ref{diag}b),
ii) the passive exchange three-body separable diagram (Fig.~\ref{diag}c),
and iii)central chain diagrams with and without exchanges
(Figs.~\ref{diag}d-g.)
Terms in these diagrams are classified as either
F-diagrams, in which the gradients in the interaction operate
on passive correlations $F_{il}$, and K-diagrams, in which
the gradients act on interacting correlations $F_{ij}$ and
the uncorrelated wavefunction $\Phi$.
The direct three-body separable diagram (\ref{diag}a) , has the general form:
\beqa
\frac{1}{A\Omega^2}
\sum_{{\bf k}_i,{\bf k}_j,{\bf k}_l}
\sum_{p,m,p',q,q'}\int
\Phi^{\ast}_3(i,j,l)
C[\frac{1}{4}\left\{f^p_{ij}O^p_{ij},f^q_{il}O^q_{il}\right\}
v^m_{ij}O^m_{ij}
\left\{f^{p'}_{ij}O^{p'}_{ij},f^{q'}_{il}O^{q'}_{il}\right\} \nonumber \\
-(f^p_{ij}O^p_{ij}v^m_{ij}O^m_{ij}f^{p'}_{ij}O^{p'}_{ij})
 (f^q_{il}O^q_{il}f^{q'}_{il}O^{q'}_{il})]
 \Phi_3(i,j,l)d^3r_{ij}d^3r_{il}, 
\eeqa
with
\beq
\Phi_3(i,j,l)=e^{i({\bf k}_i\cdot {\bf r}_i+
 {\bf k}_j\cdot {\bf r}_j+{\bf k}_l\cdot {\bf r}_l)},
\eeq
where C[\ ] represents the so-called c-part, or the $\boldsigma,
\boldtau$ -independent part of the operator product,
as described in Reference \cite{pw}. Also, the $O_{ij}$
in the separated part of the expression do not operate on
the $f_{il}O_{il}$.
The expressions for the interacting and passive exchange
diagrams (Figs.~\ref{diag}a and b) are obtained by replacing the
uncorrelated wavefunction $\Phi^\ast_3(i,j,l)$ with either
$\Phi^\ast_3(j,i,l)$ or $\Phi^\ast_3(l,j,i)$, and inserting the
appropriate exchange operators ($\frac{1}{4} \sum_{n=1,4}O^n_{ij}$
or $\frac{1}{4}\sum_{n=1,4}O^n_{il}$) to the far left of the
operator product.

The three-body separable $L^2$ diagrams with $m = 9-12$
(Figs.~\ref{diag}a-c) have large
F-parts, and smaller K-parts which we neglect. For the diagram 
presented in Fig.~\ref{diag}a we include terms with 
$p,p'=1-6$ and $q,q'=1-6$. The dominant contribution
comes from terms with $p=1$ and $q,q'$ representing either
central or tensor correlations.
The contribution from the diagrams presented in Figs.~\ref{diag}b and 
\ref{diag}c is somewhat smaller, and we only consider terms with
$p,p'=1,2$ and $q,q'=1-6$.
Contributions to the energy of SNM at $\rho=0.28$ fm$^{-3}$ for
both the U14 and A18 models, without three-body interactions
are listed in Table~\ref{diags}.

The $\ols^2$ in $O^{m=13,14}_{ij}$ can be decomposed as:
\beq
\ols^2 = -\frac{1}{2}\ols +\frac{1}{2}L^2 +\frac{1}{6}\oss L^2
+\frac{1}{6}\alpha_{ij}(L,L),
\eeq
where:
\beq
\alpha_{ij}(L,L)=3\boldsigma_i\cdot{\bf L}\boldsigma_j
\cdot{\bf L}-\oss L^2.
\eeq
The $\ols$ part of $v^{m=13,14}_{ij}$ is treated along with the
$v^{m=7,8}_{ij}$, as described by Lagaris \cite{lp}, and includes
all the diagrammatic terms therein. The $L^2$ parts
are calculated along with the $v^{m=9-12}_{ij}$, as described above.
We also include selected $\alpha_{ij}$ diagrams,
which are expected to make the most significant contribution
to the energy. These contain direct F-diagrams (Fig.~\ref{diag}a) with:
$p'q'pq= cctc,tccc,ttct,cttt,tctc$, 
and the exchange F-diagrams (Fig.~\ref{diag}b) with
$p'q'pq= cctc,tccc$, where it is understood 
that `$t$' represents  all tensor correlations with and
without $\boldtau_i\cdot\boldtau_j$ or $\boldtau_i\cdot\boldtau_l$
factors.

The central chain diagrams shown in Figs.~\ref{diag}d-g
make a modest contribution to the energy. 
The K-diagrams are obtained by inserting appropriate chain functions
into the two-body  integrals. Three central chain functions,
$G^c_{xx'=dd,de,cc}$ are defined in Reference~\cite{pw}, where
$xx'$ denote the exchange character of the two interacting vertices
of the diagram. Each of these may be either direct ($x=d$),
exchanged only with particles in the chain ($x=e$) or part of
a circular exchange involving both interacting particles ($xx'=cc$).
When considering only static interactions and correlations,
the $G^c_{xx'}$ can be directly inserted into the two-body
integrals. However, the gradients associated with
momentum-dependent interactions may operate on the Slater functions
in the de and cc chains, 
thus yielding somewhat more complicated expressions.

The direct central chain diagram (Fig.~\ref{diag}d) has neither
interacting particle exchanged. The K-part of this diagram 
is obtained in the same manner as static diagrams,
namely by dressing the two-body direct diagram with
$\left[e^{G_{dd}^c} -1\right]$.

In order to calculate the de and cc diagrams
(Figs.~\ref{diag}f and \ref{diag}g), the operator product 
$f^p_{ij}O^p_{ij}v^m_{ij}O^m_{ij}f^{p'}_{ij}O^{p'}_{ij}$ is written
in powers of $L^2$. The $L^0$ part does not contain gradients,
and the associated de and ee chain diagrams
are thus obtained as in the static case by dressing the
direct two-body diagram with:
\beq
e^{G_{dd}^c}
\left( 2G^c_{de}+ G^c_{de}G^c_{de} +G^c_{ee}\right). 
\eeq
The $L^0$ cc chain diagrams (Figs. \ref{diag}e and \ref{diag}g)
are calculated by replacing $l^2_{ij}/s$ in
the two-body exchange diagram by:
\beq
\frac{1}{s}e^{G_{dd}^c}
\left(l_{ij} +sG^c_{cc}\right)^2-\frac{l^2_{ij}}{s} \ .
\eeq
Here, $s$ is the degeneracy, and $l_{ij}\equiv l(k_fr_{ij})$
is the Slater function.
The K-contribution of terms in the de and cc diagrams
containing $L^2$ operators are calculated using the chain functions
$G^{L^2}_{de}$ and $G^{L^2}_{cc}$ defined as:
\beqa
G^{L^2}_{de} &=& -\frac{\rho r^2_{ij}}{s}
\int d^3r_l F^c_{lj} l_{il}\tilde{l}_{il} \\
G^{L^2}_{cc} &=& -\frac{\rho r^2_{ij}}{s^2}l_{ij}
\int d^3r_l F^c_{lj} l_{lj}\tilde{l}_{il}
\eeqa
with
\beq
\tilde{l}_{il}\equiv
\frac{1}{4}\cos^2\theta_i\left(l''_{il}
-\frac{l'_{il}}{r_{il}}\right)
-\frac{1}{4}\left(l''_{il}+\frac{l'_{il}}{r_{il}}\right)
+\cos\theta_i\frac{l'_{il}}{r_{ij}} \ .
\eeq
The $L^2$ chain functions result from the $L^2_{ij}$ acting on the
Fermi gas part of the wavefunctions. These
depend on $r_i$ and $r_j$ due to the exchanges at the
interacting vertices of the chain diagrams.

The de contribution of the $L^2$ part is obtained by dressing
the two-body direct diagram with:
\beq
e^{G_{dd}^c}
\left( G^c_{de}+ 2(5/k^2_fr^2_{ij})G^{L^2}_{de}\right), 
\eeq
while the cc contribution is calculated by replacing 
$r_{ij}l_{ij}l'_{ij}/s$ in the two-body exchange diagram by:
\beq
e^{G_{dd}^c}
\left(\frac{r_{ij}l_{ij}l'_{ij}}{s}
-r_{ij}l'_{ij}G^c_{cc}+2G^{L^2}_{cc}\right)
-\frac{r_{ij}l_{ij}l'_{ij}}{s} \ .
\eeq

The present calculation of $\langle v \rangle$-MB-q
includes terms involving only the static correlations,
since they yield the dominant contribution to the energy.
Without $\ols$ correlations, 
the $f^p_{ij}O^p_{ij}v^m_{ij}O^m_{ij}f^{p'}_{ij}O^{p'}_{ij}$
operator product does not contain terms with $L^{n>2}$, and the above 
calculation of central chain K-diagrams is complete.

\section{PERTURBATIVE CORRECTIONS}

We wish to calculate the two-body cluster using a better variational
wavefunction. The standard two-body variational wavefunction has the form:
\beq
\Psi_{2b}=(\sum_{p=1,8} f^p(r_{ij})O^p_{ij})\Phi_{2b},
\eeq
where $\Phi_{2b}$ is a plane-wave Slater determinant. The
$f^p$ are determined from the E-L equations obtained by 
minimizing the total two-body cluster energy
\beq
C_2=\frac{1}{A}\sum_{m,n}\langle\Psi_{2b}|H-\frac{\hbar^2}{m}k^2_{mn}
|\Psi_{2b}\rangle 
   =\frac{1}{A}\sum_{{\bf k}_m,{\bf k}_n}
   \sum_{\boldsigma\!\!_m,\boldsigma\!\!_n}
   \sum_{\boldtau\!\!_m,\boldtau\!\!_n}
\langle\Psi_{2b}|H-\frac{\hbar^2}{m}k^2_{mn}
|\Psi_{2b}\rangle ,
\eeq
obtained after summing over ${\bf k}_m$, ${\bf k}_n$, $\boldsigma\!_m$, 
$\boldsigma\!_n$, $\boldtau\!_m$ and $\boldtau\!_n$
at chosen values of $d_c$, $d_t$ and $\alpha$.
We propose instead to minimize $C_2(k_m,k_n)$ separately in
each partial wave, thus obtaining momentum- and channel-dependent
correlations $f(l,S,J,k)$.
The perturbative correction to the energy in an $l,S,J$ channel
is then given by:
\beq
\delta E_2(l,S,J)=\sum_{{\bf k}_m,{\bf k}_n}
[C_2(l,S,J;k_m,k_n)-\bar{C}_2(l,S,J;k_m,k_n)],
\eeq
where $C_2$ is obtained from the $f(l,S,J,k)$ and
$\bar{C}_2$ is calculated with the operator $f^pO^p$.

The uncorrelated two-body wavefunction, in relative coordinates, can
be expanded as:
\beqa
e^{i kz} |S,M\rangle &=& \delta_{S0}\delta_{M0} 
\sum_J \sqrt{4\pi(2J+1)}\ i^J  j_{_J}(kr) \ {\cal Y}^0_{J0J}\nonumber \\
&+&\delta_{S1} \sum_J \sqrt{4\pi(2J+1)}\ i^J  j_{_J}(kr) 
\ {\cal Y}^M_{J1J}\langle J1JM|J01M\rangle\nonumber \\
&+&\delta_{S1}\sum_{J\geq 1} \sqrt{4\pi} \ i^{J_-}
[\xi^M_{J_-}(r){\cal Y}^M_{J_-1J}+\xi^M_{J_+}(r){\cal Y}^M_{J_+1J}],
\eeqa
where $J_\pm = J\pm 1$, ${\cal Y}^M_{lSJ}$ are spin-angle functions,
$\langle lSJM|lm_lSm_s \rangle$ are Clebsch-Gordon coefficients
and:
\beqa
\xi^{M}_{J_-}(r) &=& \ \ \sqrt{(2J_- +1)}\ 
\langle J_- 1JM| J_-01M\rangle \ j_{_{J_-}}(kr)  \nonumber \\
\xi^{M}_{J_+}(r) &=& -\sqrt{(2J_+ +1)}\ 
\langle J_+ 1JM| J_+01M\rangle \ j_{_{J_+}}(kr) \ . 
\eeqa
The correlated two-body wavefunction can thus be written:
\beqa
\Psi(r,k,S,M) &=& \sum_J\sqrt{4\pi(2J+1)}\ i^J
[R_{J,S=0}(r)\ {\cal Y}^0_{J0J} + R_{J,S=1}(r)
\ {\cal Y}^M_{J1J}\langle J1JM|J01M\rangle]\nonumber \\
&+&\sum_{J\geq 1}\sqrt{4\pi} \ i^{J_-}
[R^M_{J_-}(r){\cal Y}^M_{J_-1J}+R^M_{J_+}(r){\cal Y}^M_{J_+1J}].
\eeqa
The wavefunction $u_{J=l,S=0,1}=rR_{J,S}$ in uncoupled channels
satisfies the E-L equation:
\beq
-\hovm u^{\prime\prime}_{J,S} +[\hovm\frac{J(J+1)}{r^2}+\bar{v}_{J,S}
-\lambda_{J,S}(k)]u_{J,S}=\hovm k^2 u_{J,S},
\eeq
and is normalized such that:
\beq
u_{J,S}(r=d_x)=rj_{_J}(kr).
\eeq
The $\lambda_{J,S}(k)$ are constants, which are varied to match 
the boundary conditions:
\beq
u^\prime_{J,S}(r=d_x)=j_{_J}(kr)+rj^\prime_{_J}(kr),
\eeq
where $d_x=d_c$ in the spin-singlet state, and $d_x=d_t$ in the
uncoupled spin-triplet state, which is affected by the tensor
correlations.

In the coupled channels, $J=l\pm1, S=1$, we obtain a pair
of coupled equations:
\beqa
-\hovm u^{\prime\prime} +[\hovm\frac{J_-(J_-+1)}{r^2}
+\bar{v}_{J_-,1} -\lambda^c(k)]u
+\frac{6\sqrt{J(J+1)}}{2J+1}(\bar{v}^t_{J,1}-\lambda^t(k))\omega
=\hovm k^2 u  \\
-\hovm \omega^{\prime\prime}+[\hovm\frac{J_+(J_++1)}{r^2}
+\bar{v}_{J_+,1} -\lambda^c(k)]\omega
+\frac{6\sqrt{J(J+1)}}{2J+1}(\bar{v}^t_{J,1}-\lambda^t(k))u
=\hovm k^2 \omega.
\eeqa
These equations have two sets of solutions, denoted by
$(u_-,\omega_-)$ and $(u_+,\omega_+)$, with boundary conditions:
\beqa
&u_-(r=d_x)&=rj_{_{J_-}}(kr) \ ; \ \omega_-(r=d_t)=0, \nonumber \\
&u^\prime_-(r=d_x)&=j_{_{J_-}}(kr)+rj^\prime_{_{J_-}}(kr) \ ; \
\omega^\prime_-(r=d_t)=0,
\eeqa
where $d_x=d_c$ for $J=1,l=J_-=0$, $^3S_1$ state, and $d_t$
otherwise, and
\beqa
&u_+(r=d_t)&=0 \ ; \ \omega_+(r=d_t)=rj_{J_+}(kr), \nonumber \\
&u^\prime_+(r=d_t)&=0 \ ; 
\ \omega^\prime_+(r=d_t)=j_{_{J_+}}(kr)+rj^\prime_{_{J_+}}(kr).
\eeqa
The $\lambda^c(k)$, is adjusted to match the derivative boundary condition
on the dominant wave ($u_-$ or $\omega_+$), while $\lambda^t(k)$
is varied to match the zero derivative condition on the
secondary wave ($u_+$ and $\omega_-$).
Thus, $\lambda^c(k)$ and $\lambda^t(k)$ depend on  $J$ and the $l$
of the dominant wave.

The $R^M_{J_\pm}$ can be expressed as superpositions of
the two solutions, which match the boundary condition:
$R^M_{J_\pm}(r\rightarrow\infty)=\xi^M_{J_\pm}(r\rightarrow\infty)$.
Evaluating the Clebsch-Gordon coefficients in $\xi^M_{J_\pm}$,
we find:
\beqa
R^{M=\pm 1}_{J_-} &=& \frac{1}{r}\left[ u_-\sqrt{\frac{J+1}{2}}-u_+
\sqrt{\frac{J}{2}} \ \right] \nonumber \\
R^{M=\pm 1}_{J_+} &=& \frac{1}{r}\left[
\omega_-\sqrt{\frac{J+1}{2}}-\omega_+ \sqrt{\frac{J}{2}} 
\ \right] \nonumber \\
R^{M=0}_{J_-} &=&\frac{1}{r}\ \left[ u_-\sqrt{J} +u_+\sqrt{J+1}
\ \right] \nonumber \\
R^{M=0}_{J_-} &=&\frac{1}{r}\ \left[\omega_-\sqrt{J} +\omega_+\sqrt{J+1}
\ \right]. 
\eeqa

With the wavefunction in hand, we can calculate the two-body cluster
energy, and thereby determine $\delta E_2(l,S,J)$.
The differential equations for the wavefunctions are solved at
several values of $k$ on a grid from 0 to $k_f$. The expressions
for $C_2(l,S,J,k_m,k_n)$ are then integrated over the Fermi sea.
In order to insure that the correlations functions, $f$,
in coupled channels are positive, the solutions
are matched to the asymptotic forms at the first node of the
bessel function, rather than at the healing distance ($d_c$ or $d_t$)
for values of $k$  large enough such that the node occurs within the 
healing distance.

The momentum-dependence of $f(l,S,J,k)$ is not very large, however the
dependence on $l,S,J$ channels is significant.
We find that no additional attraction is obtained in the $^3S_1-^3\!D_1$
channel, and very little comes from the $^1S_0$ and other singlet
channels as can be seen in Table~\ref{de2}. This presumably 
indicates that the parameters $d_c$, $d_t$ and $\alpha$ are
optimum for those channels. The bulk of the additional 
attraction due to channel- and momentum-dependent
correlations is in the $^3P_2-^3\!F_2$ and $^3P_1$ channels, with somewhat
less coming from the $^3D_2$ channel.
The present $f^pO^p$ correlation operator, with $d^p$ chosen according
to Eqs.~3.8 and 3.9, is probably inadequate to simultaneously describe
correlations in $S$- and $P$-waves accurately. However, the $\delta E_2$
correction has little effect on the critical density of the predicted
phase transition.

\newpage

\begin{figure}
\caption{$E(\rho)$ of SNM calculated using A18+UIX. Included for 
comparison are previous calculations of $E(\rho)$ using A14+UVII (WFF), 
and U-DDI (FP). The two sets of variational minima obtained at
$\rho >0.28$ fm$^{-3}$ are labeled LDP and HDP.}
\label{esnm}
\end{figure}

\begin{figure}
\caption{$E(\rho)$ 
of PNM calculated using A18+UIX. Included for comparison are
previous calculations of $E(\rho)$ using A14+UVII (WFF), and U-DDI (FP).
The two sets of variational minima obtained at
$\rho >0.16$ fm$^{-3}$ are labeled LDP and HDP.}
\label{epnm}
\end{figure}

\begin{figure}
\caption{Constrained energies:
$E_c(\rho,d_t/r_0)= \frac{1}{2}(E_{PB}+E_{JF})
+ \Lambda \left[ (I_c-1)^2 +(\frac{1}{3}I_\tau+1)^2 \right]$,
of SNM using A18+UIX. $\Lambda$ was set to 1000 MeV, in order to
keep the integrals of the two-body densities $I_c$ and $I_\tau$
within 5\% of their exact values of 1 and $-3$ during minimization.}
\label{edtsnm}
\end{figure}

\begin{figure}
\caption{Constrained energies:
$E_c(\rho,d_t/r_0)= \frac{1}{2}(E_{PB}+E_{JF})+ \Lambda (I_c-1)^2 $ of PNM,
using A18+UIX. $\Lambda$ was set to 1000 MeV.}
\label{edtpnm}
\end{figure}

\begin{figure}
\caption{$\rho^{p=1-6}_2$ for SNM at $\rho=0.16$ fm$^{-3}$}
\label{snmrho16}
\end{figure}

\begin{figure}
\caption{$\rho^{(2)}_{T,S,M=0,1,0}(r;\theta=0,\pi/2)$ for
SNM at $\rho=0.16$ fm$^{-3}$, and for light nuclei scaled to
match the maximum value of the SNM distribution.}
\label{snmrts01}
\end{figure}

\begin{figure}
\caption{$\rho^{(2)}_{T,S,M=1,0,0}(r)$ for SNM at $\rho=0.16$ fm$^{-3}$,
and for light nuclei scaled to match the maximum value of the SNM
distribution.}
\label{snmrts10}
\end{figure}

\begin{figure}
\caption{Two-body densities $\rho^{(2)}_{T,S,M}$
for SNM at $\rho=0.36$ fm$^{-3}$.
The full and dashed lines represent results for the LDP and HDP, respectively}
\label{snmrho36}
\end{figure}

\begin{figure}
\caption{Two-body densities $\rho^{(2)}_{S,M}$ for PNM
at $\rho=0.2$ fm$^{-3}$.
The full and dashed lines represent results for the LDP and HDP, respectively}
\label{pnm111020}
\end{figure}

\begin{figure}
\caption{$v^{\sigma\tau}(r)$ and $\tilde{v}^\sigma(r)$
in U14, A14 and A18 models of the nucleon-nucleon interaction.}
\label{vst}
\end{figure}

\begin{figure}
\caption{$S_L(q)$ for SNM and PNM.}
\label{slq}
\end{figure}

\begin{figure}
\caption{$W_L(q)/S_L(q)$ for SNM and PNM.
$q^2/2m$ (dash-dot curve) is plotted for comparison.}
\label{wos}
\end{figure}

\begin{figure}
\caption{$\xi(q)$ for SNM and PNM.}
\label{xiq}
\end{figure}

\begin{figure}
\caption{Pion excess function, $\eta(q)$ for SNM and PNM.}
\label{etaq}
\end{figure}

\begin{figure}
\caption{(a)3-body separable direct, (b)3-body separable, interacting
exchange, (c)3-body separable, passive exchange,
(d) direct central chain, (e) interacting exchange central chain
(f) pair exchange central chain and
(g) circular exchange central chain diagrams.}
\label{diag}
\end{figure}

\newpage

\begin{table}
\caption{Contributions to the Energy of SNM in MeV}
\begin{tabular}{lrrrrrr}
            &U14    &U14    &A18    &A18    &A18+IX &A18+IX \\
\hline

$k_f$ (fm$^{-1}$)
	    &   1.33&   1.60&   1.33&   1.60&   1.33&   1.60\\
$\rho \ \,$ (fm$^{-3}$)
	    &    .16&    .28&    .16&    .28&    .16&    .28\\
$d_c \,$ (fm)
	    &   2.15&   1.79&   2.13&   2.08&   1.80&   1.67\\
$d_t \,$ (fm)
	    &   3.44&   2.86&   5.67&   4.76&   4.81&   3.81\\
$\alpha$    &    .80&    .80&    .65&    .61&    .80&    .90\\
\hline
$\langle T \rangle$-1B 
	    &  22.01&  31.85&  22.01&  31.85&  22.01&  31.85\\
$\langle v \rangle$-2B-s
	    & -56.40& -75.21& -66.17&-100.68& -66.41&-105.34\\
$\langle T \rangle$-2B-s 
	    &  16.08&  21.72&  19.01&  28.51&  20.25&  33.47\\
$\langle v \rangle$-2B-b 
	    &  -3.23&  -6.37&  -2.29&  -5.02&  -2.38&  -5.76\\
$\langle T \rangle$-2B-b
	    &    .80&   1.50&    .54&   1.10&    .62&   1.58\\
$\langle v \rangle$-2B-q 
	    &   1.11&   3.56&   4.31&  11.39&   4.46&  12.44\\
$\langle v+T \rangle$-2B
	    & -41.64& -54.80& -44.61& -64.71& -43.46& -63.61\\
$\langle v+T \rangle$-MB-s  
	    &   4.76&   2.91&   6.47&   9.31&   5.50&   8.47\\
$\langle v+T \rangle$-MB-b  
	    &   -.21&  -1.02&   -.28&  -1.01&   -.22&   -.72\\
$\langle v \rangle$-MB-q 
	    &   1.61&   3.91&   3.00&   8.54&   3.38&  10.62\\
$\langle V^{2\pi} \rangle$  
	    &    .00&    .00&    .00&    .00&  -3.60&  -9.96\\
$\langle V^R \rangle$ 
	    &    .00&    .00&    .00&    .00&   6.33&  22.09\\
$\delta E_{2B}$
            &  -0.95&  -2.06&  -1.30&  -2.35&  -1.89&  -4.07\\
$\frac{1}{2}(E_{PB}+E_{JF})+\delta E_{2B}$
            & -14.42& -19.22& -14.71& -18.37& -11.96&  -5.33\\
$\frac{1}{2}(E_{PB}-E_{JF})$
            &    .30&    .27&    .55&    .82&    .60&   1.10\\

\end{tabular}
\label{tabenm}
\end{table}

\begin{table}
\caption{Contributions to $E(\rho,d_t/r_0)$ of SNM for A18+IX model in MeV}
\begin{tabular}{lrrrrr}
$\rho \ $ (fm$^{-3}$)            &    .16&    .28&    .28&    .36&    .36 \\
$d_t/r_0$                        &   4.20&   4.00&   6.00&   4.00&   6.00 \\
$d_c/d_t$                        &    .38&    .41&    .25&    .44&    .28 \\
$\alpha$                         &    .80&    .90&    .96&    .98&    .88 \\
\hline
$\langle T \rangle$-1B           &  22.11&  32.10&  32.10&  37.96&  37.96 \\
$\langle v+T \rangle$-2B         & -43.65& -62.71& -67.24& -74.37& -79.54 \\
$\langle v+T\rangle$-MB          &   8.71&  18.11&  35.29&  28.52&  43.70 \\
$\langle T \rangle$              &  42.26&  67.41&  72.28&  85.16&  85.97 \\
$\langle v \rangle$              & -55.09& -79.91& -72.13& -93.06& -83.85 \\
$\langle v^{t\tau} \rangle$      & -29.14& -45.82& -49.65& -56.94& -59.23 \\
$\langle v^{\sigma\tau} \rangle$ & -10.30& -12.85&  -6.20& -13.81&  -6.97 \\
$\langle v^c \rangle$            & -26.04& -47.27& -47.10& -61.65& -61.25 \\
$\langle V^{2\pi} \rangle$       &  -3.64& -10.72& -22.88& -18.35& -32.26 \\
$\langle V^R \rangle$            &   6.42&  22.65&  23.76&  40.69&  41.71 \\
$\delta E_{2B}$ 
				 &  -1.91&  -4.65&  -2.26&  -6.78&  -6.20 \\
$\frac{1}{2}(E_{PB}+E_{JF})+\delta E_{2B}$
                                 & -11.96&  -5.22&  -1.22&   7.67&   5.37 \\
$\frac{1}{2}(E_{PB}-E_{JF})$
                                 &    .60&   1.05&   2.30&   1.45&   2.82 \\

\end{tabular}
\label{tabedt}
\end{table}

\begin{table}
\caption{Expectation values of the Pion Exchange Interaction and Pion Excess
Operators}
\begin{tabular}{lrrr}
&$\langle v^\pi\rangle/A$, MeV& $\langle V^{2\pi}\rangle/A$, MeV&
                                $\langle\delta n^{(1)}_\pi\rangle/A$ \\
\hline
SNM, $\rho=0.16$ fm$^{-3}$           & -31.53  & -3.64 &  0.05\\
SNM, $\rho=0.36$ fm$^{-3}$, LDP      & -60.62  &-18.35 &  0.09\\
SNM, $\rho=0.36$ fm$^{-3}$, HDP      & -66.86  &-32.26 &  0.15\\
PNM, $\rho=0.16$ fm$^{-3}$           &  -0.67  &  1.23 & -0.01\\
PNM, $\rho=0.20$ fm$^{-3}$, LDP      &   0.14  &  1.90 & -0.01\\
PNM, $\rho=0.20$ fm$^{-3}$, HDP      & -18.20  & -8.67 &  0.05\\
\end{tabular}

\label{vpinpi}
\end{table}

\begin{table}
\caption{MB-q Contributions in MeV 
to SNM at $\rho=0.28$ fm$^{-3}$ using U14 and A18 NN-interactions}
\begin{tabular}{lrr}
diagram & U14 & A18 \\
\hline
2B-dir & 4.18 & 9.12\\
2B-ex  &-0.62 & 2.27\\
(a)    & 3.47 & 8.57\\
(b)   & 0.22 &-1.54\\
(c)  & 0.21 & 0.46\\
(d+f) & 0.20 & 0.49\\
(e+g)   &-0.19 & 0.56\\
\end{tabular}
\label{diags}
\end{table}

\begin{table}
\caption{Contributions to $\delta E_2$ from $l,S,J$ channels for A18+IX model
 in MeV}
\begin{tabular}{lccccccccc}
$\rho$ (fm$^{-3}$)& $^1S_0$&$^1P_1$&$^1D_2$&$^1F_3$&$^3P_0$&$^3P_1$&$^3D_2$&$^3F_3$&
$^3P_2-^3F_2$\\
\hline
0.16 (SNM LDP)& 0.00& 0.00&-0.01&0.00&-0.20&-0.80&-0.15& 0.00&-0.94\\
0.28 (SNM LDP)&-0.03& 0.00&-0.03&0.00& 0.00&-1.74&-0.24& 0.00&-2.16\\
0.36 (SNM HDP)&-0.05& 0.00&-0.05&0.00&-0.14&-2.97&-0.35& 0.00&-2.80\\
0.16 (PNM LDP)& 0.00&     &-0.01&    &-0.15&-0.62&     & 0.00&-0.70\\
0.20 (PNM LDP)& 0.00&     &-0.01&    &-0.16&-0.83&     &-0.01&-0.15\\
0.24 (PNM HDP)&-0.22&     &-0.09&    &-0.41&-1.10&     & 0.00&-0.20\\
\end{tabular}
\label{de2}
\end{table}

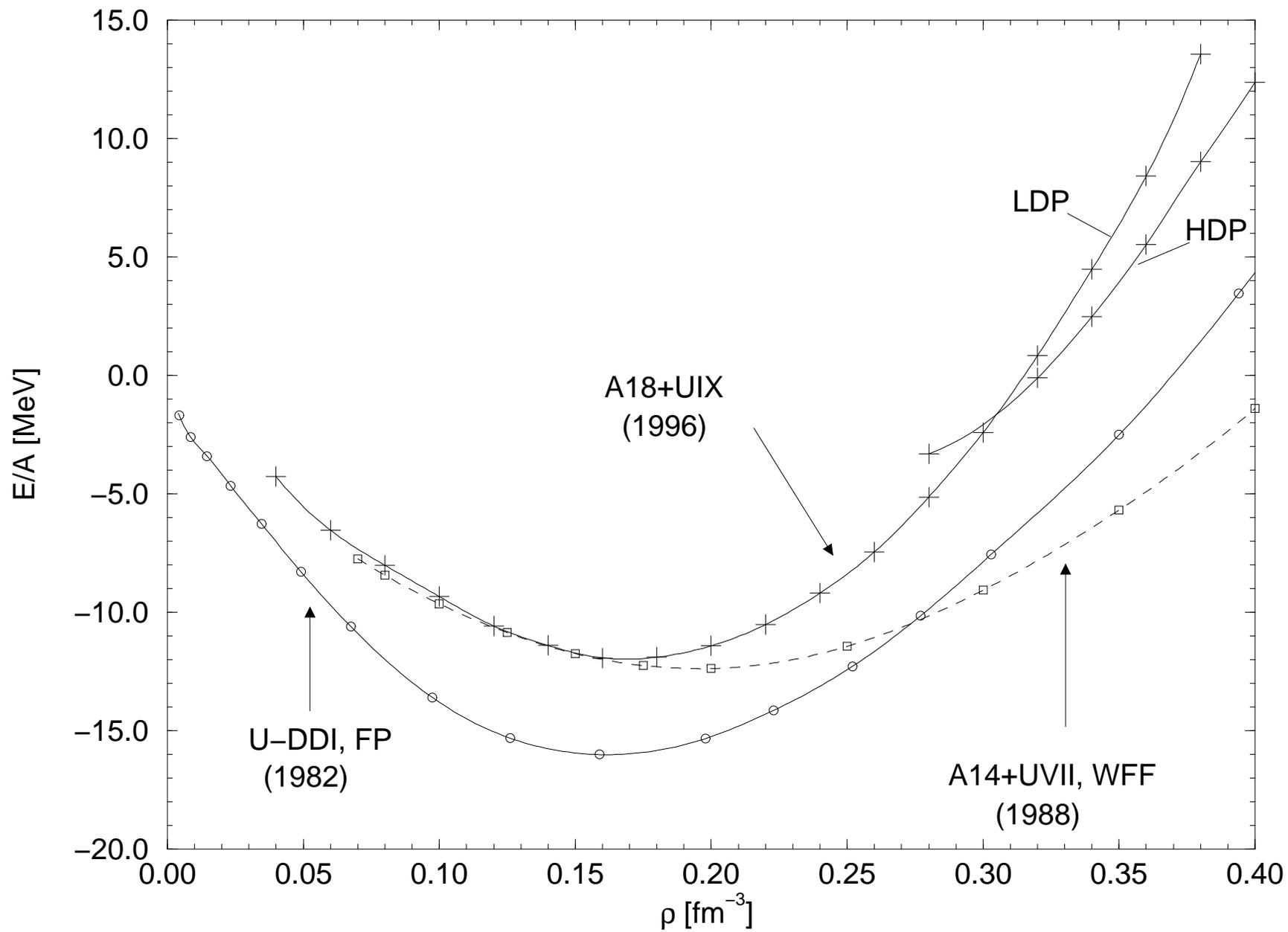
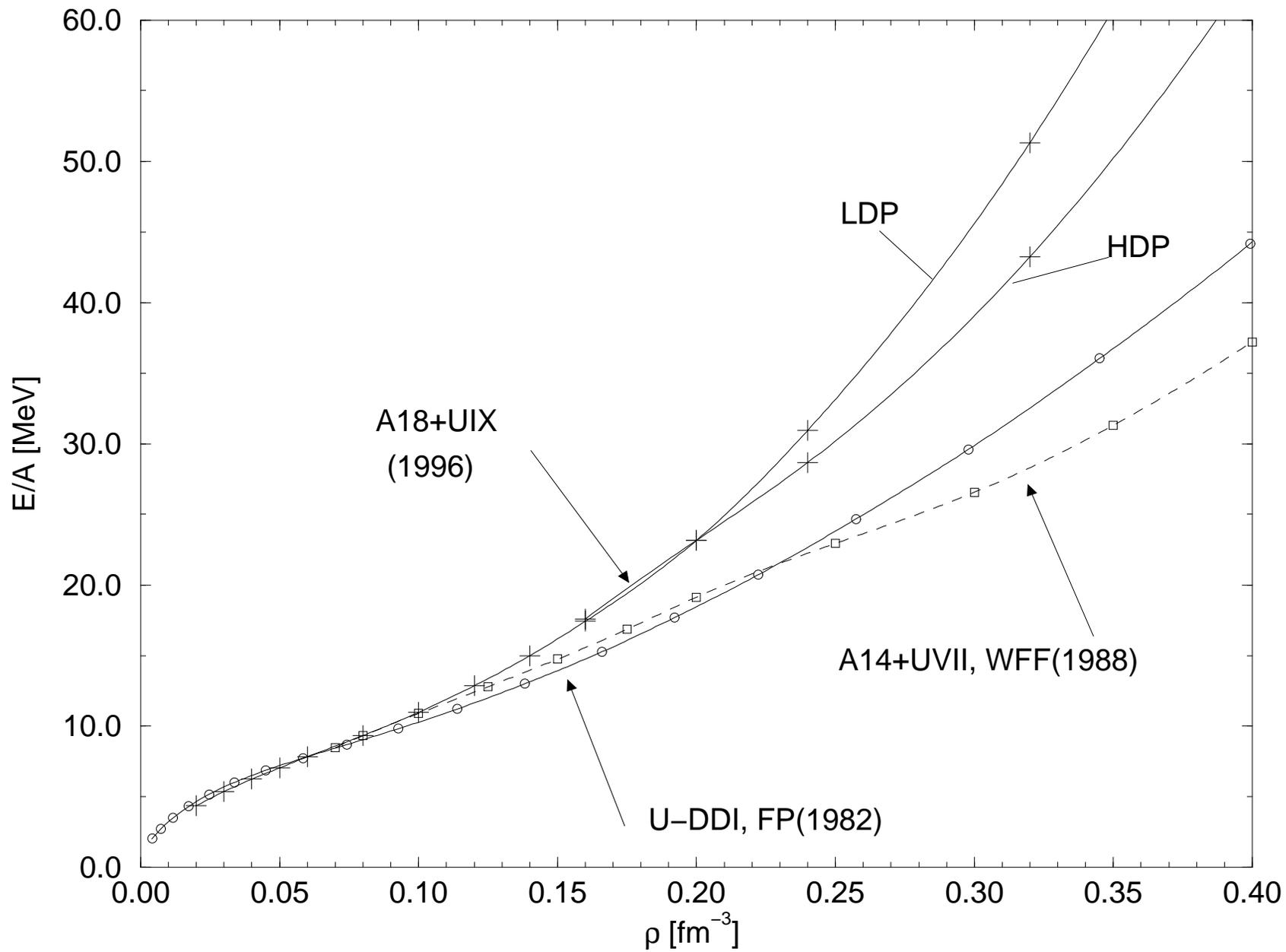
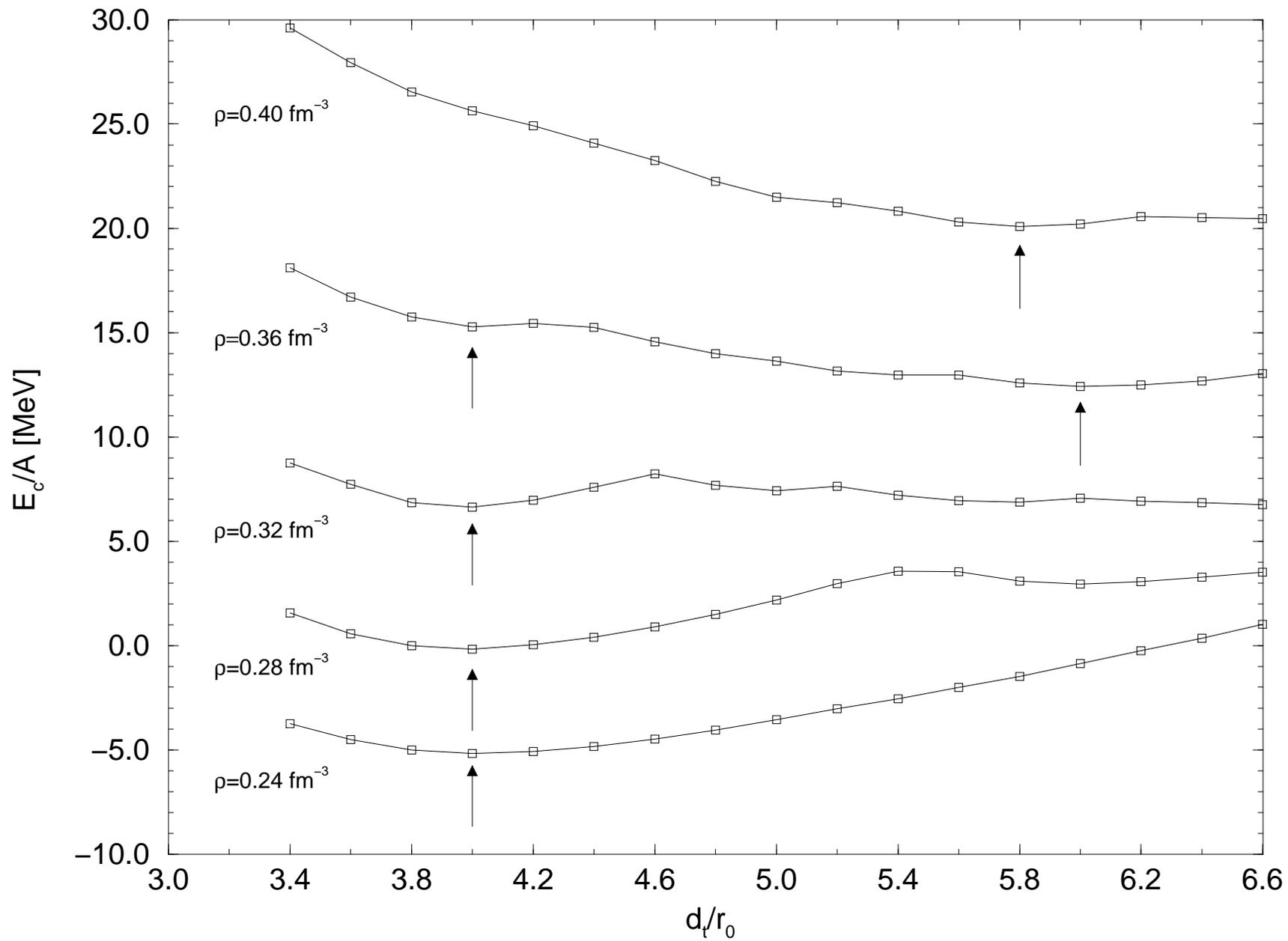
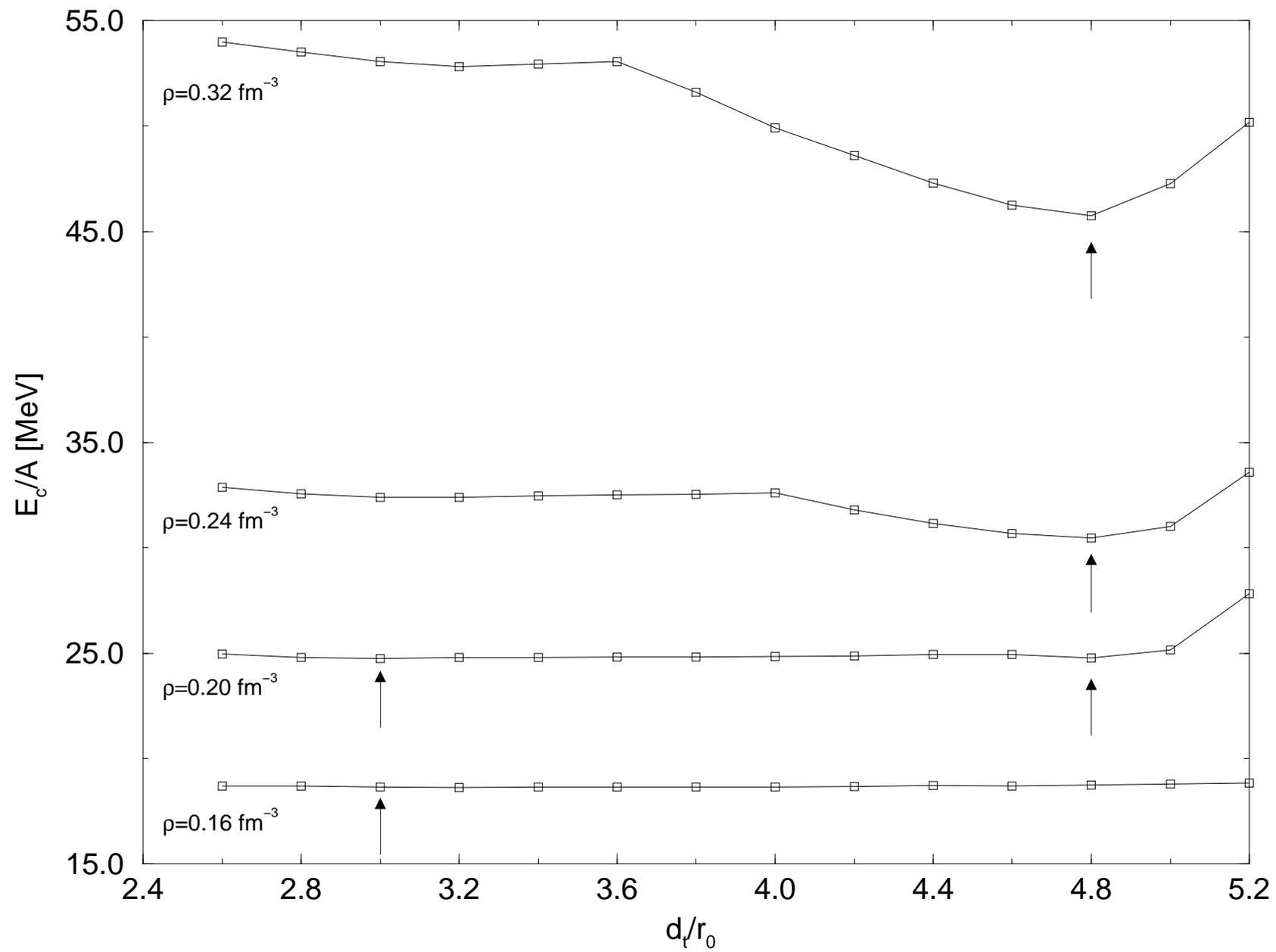
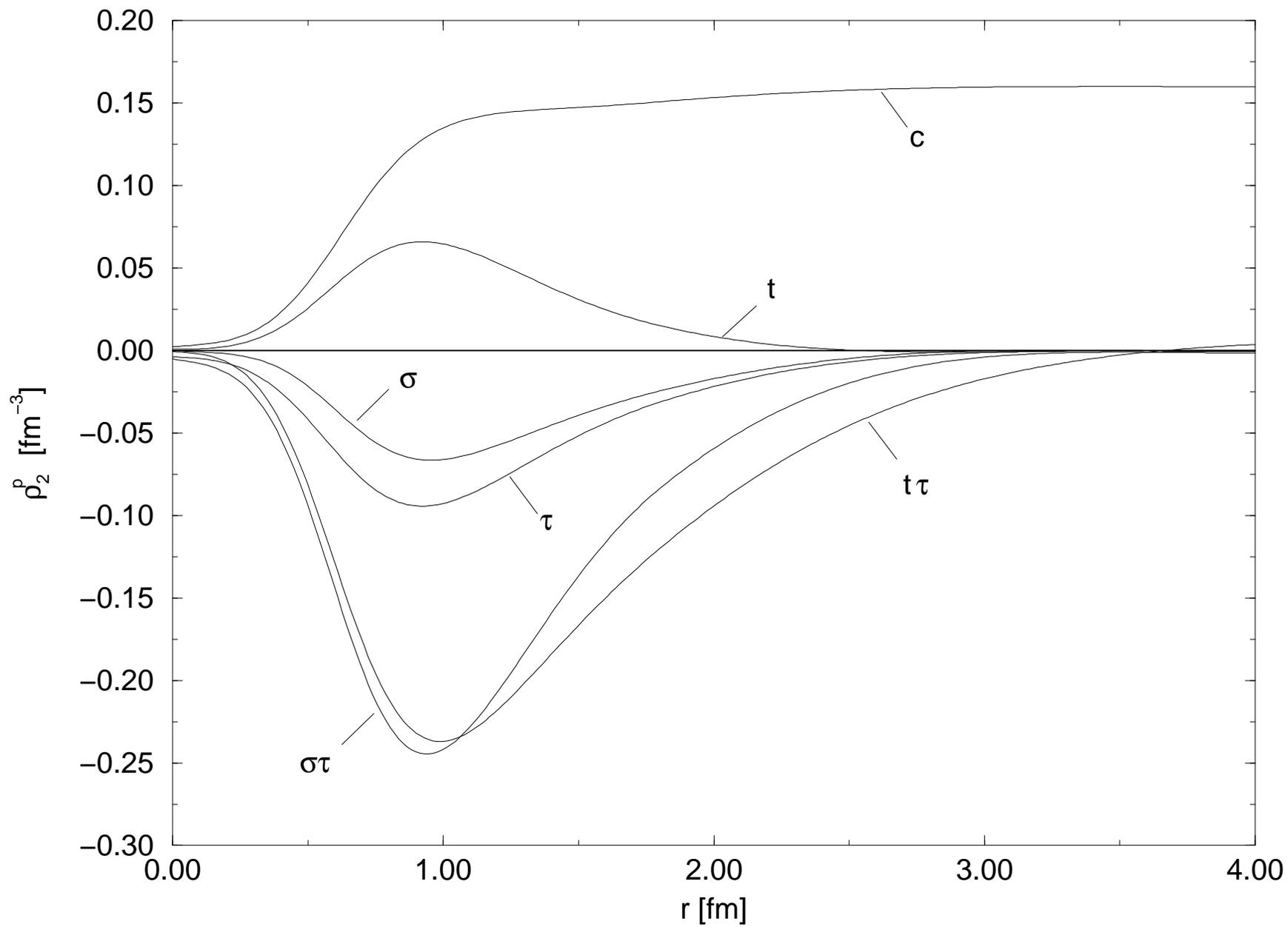
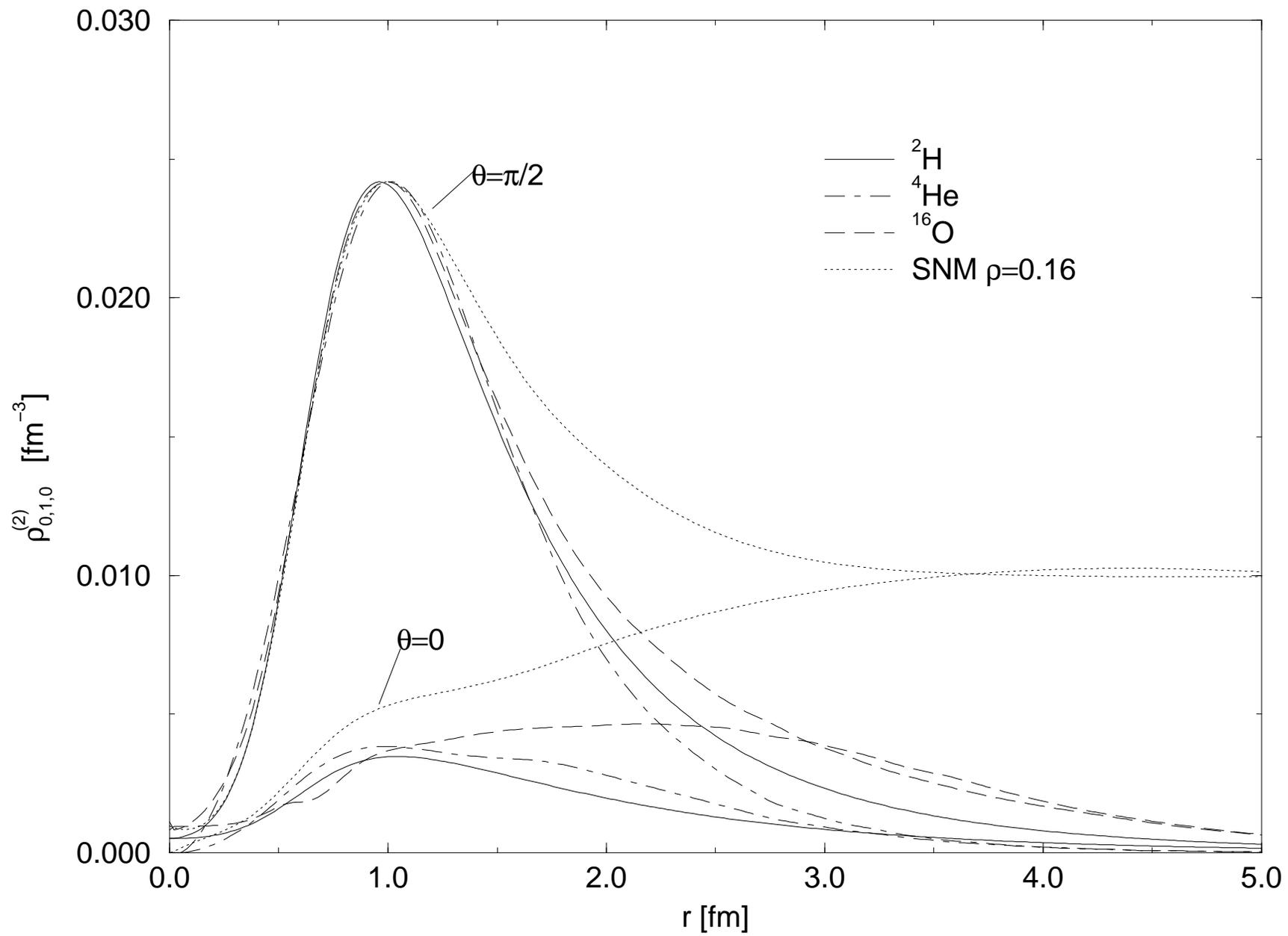
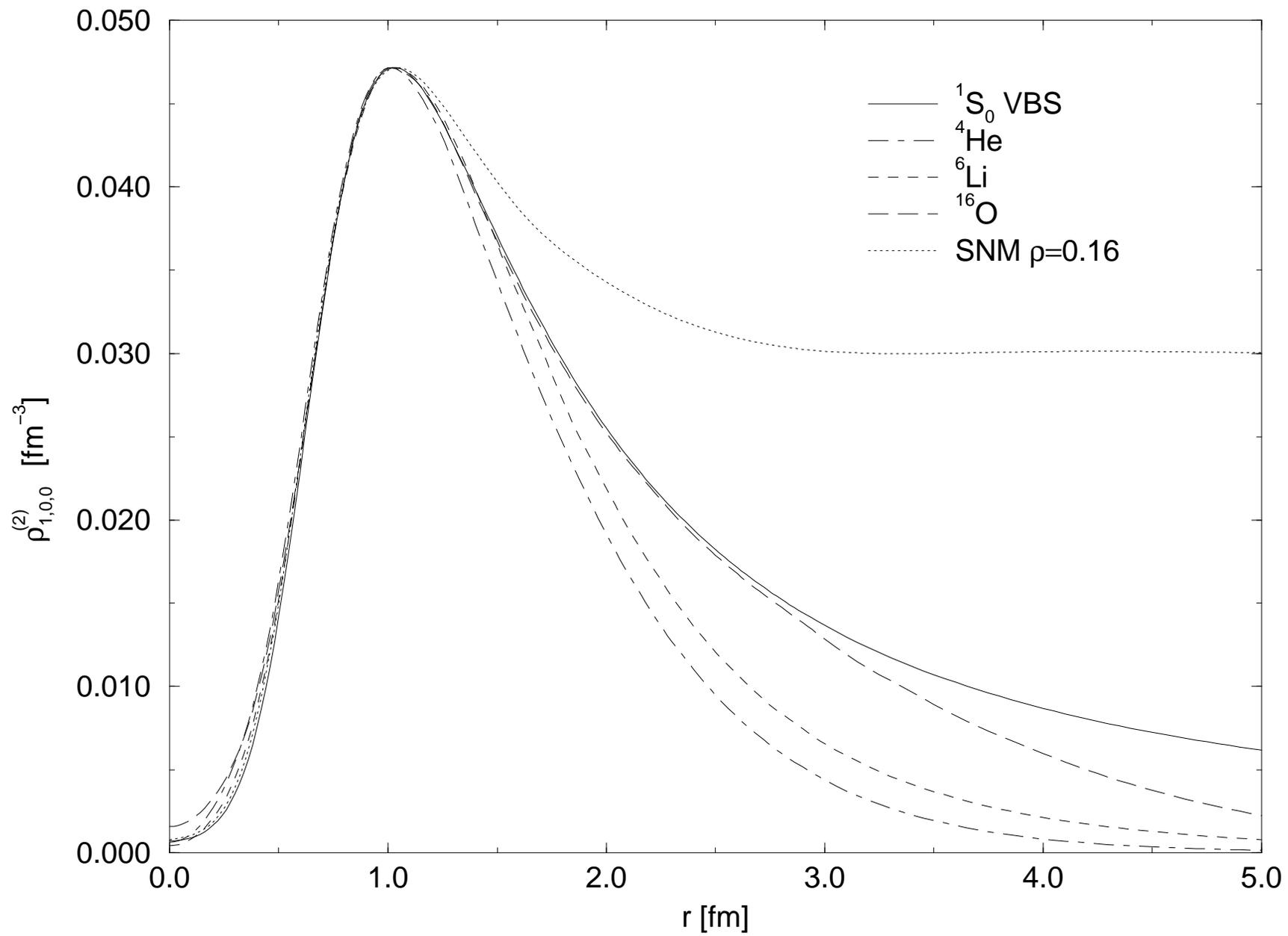
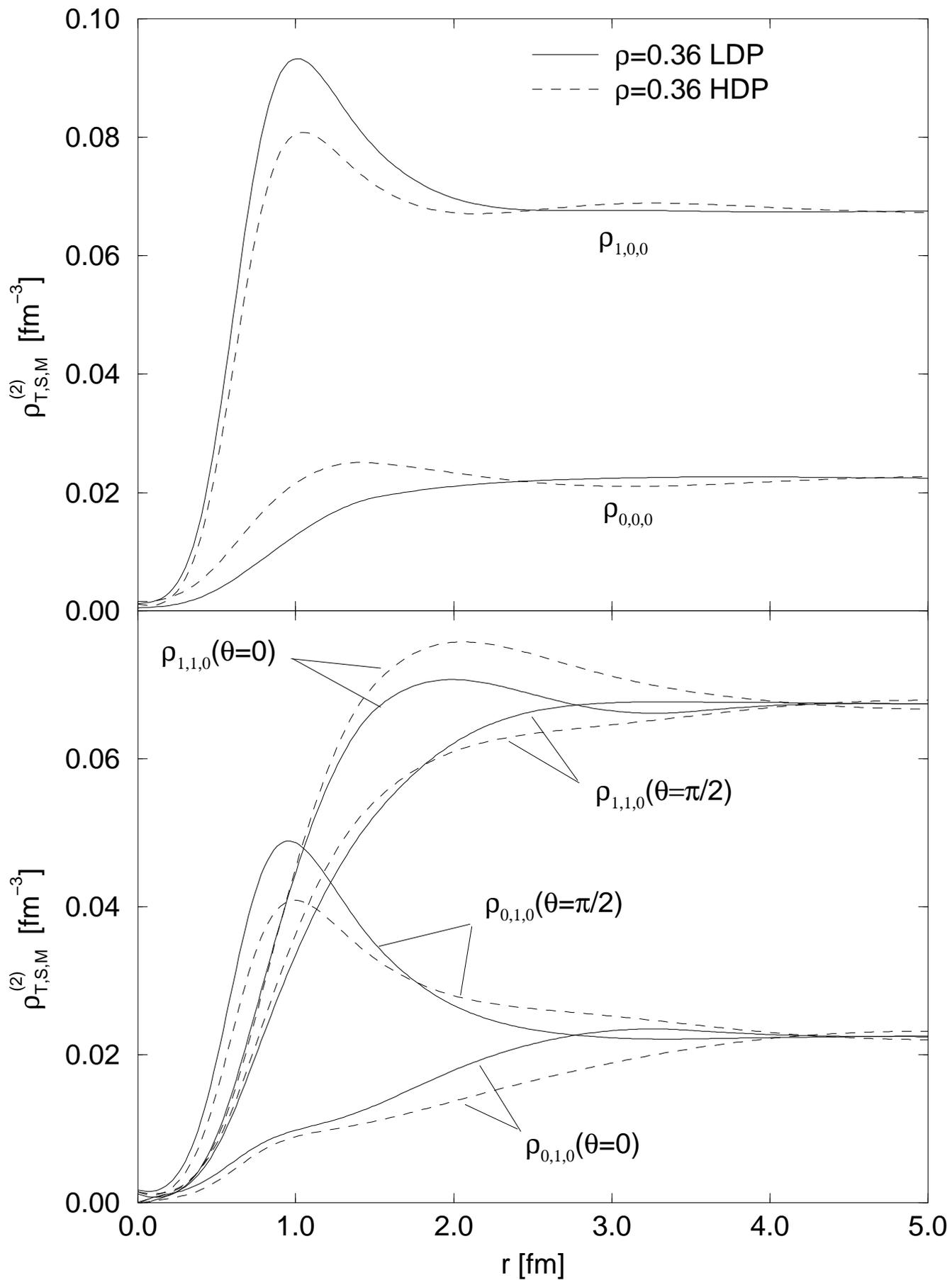
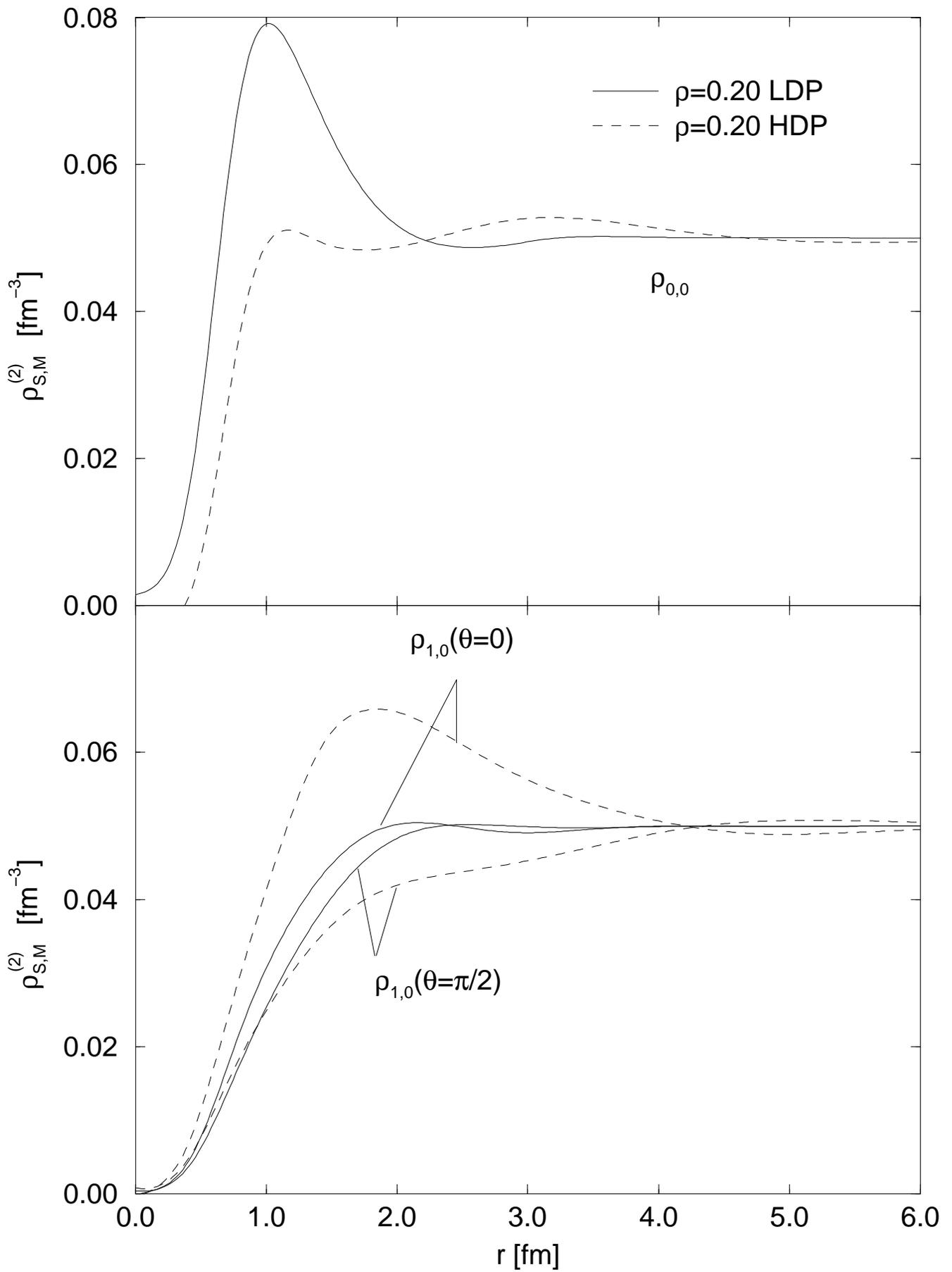
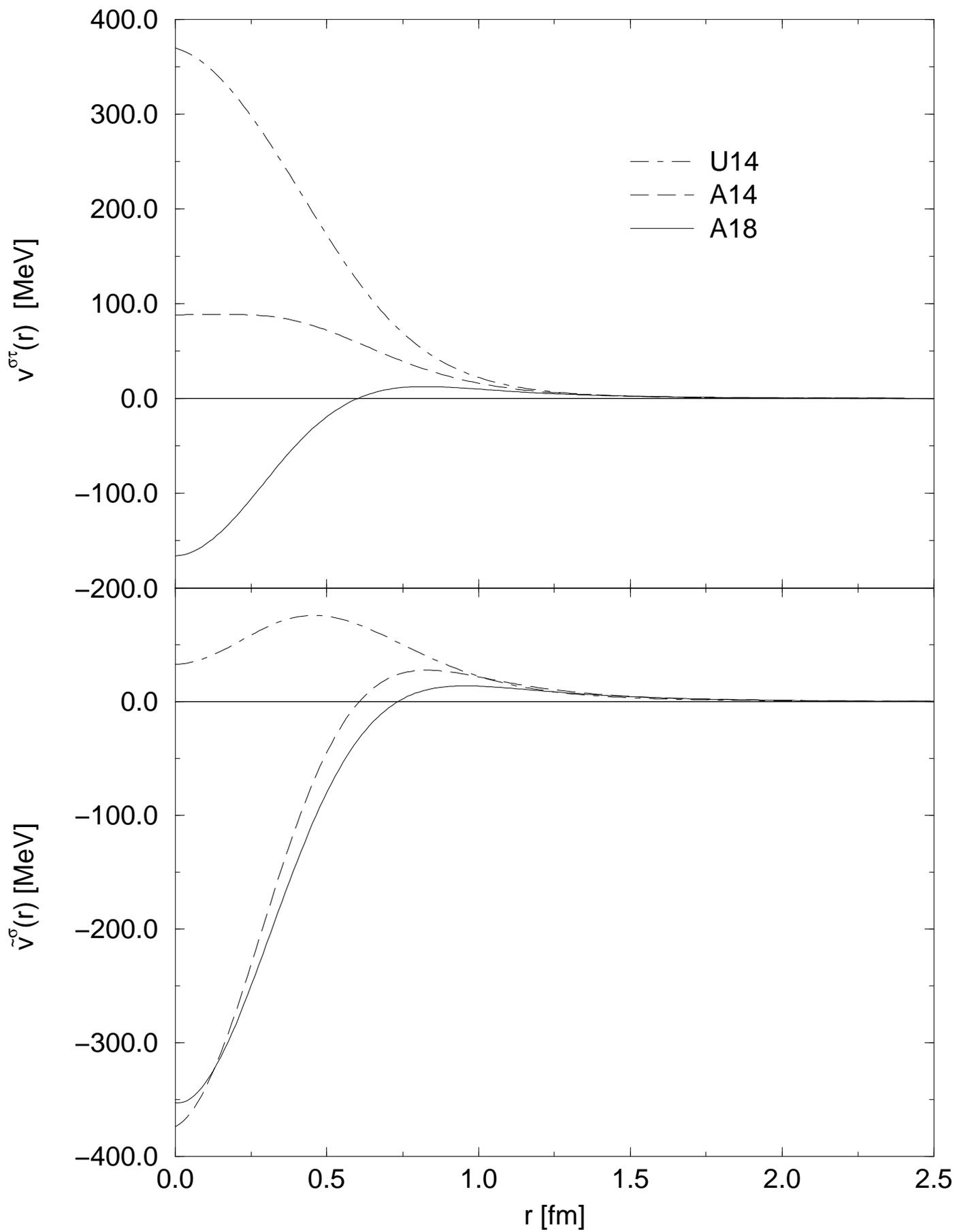
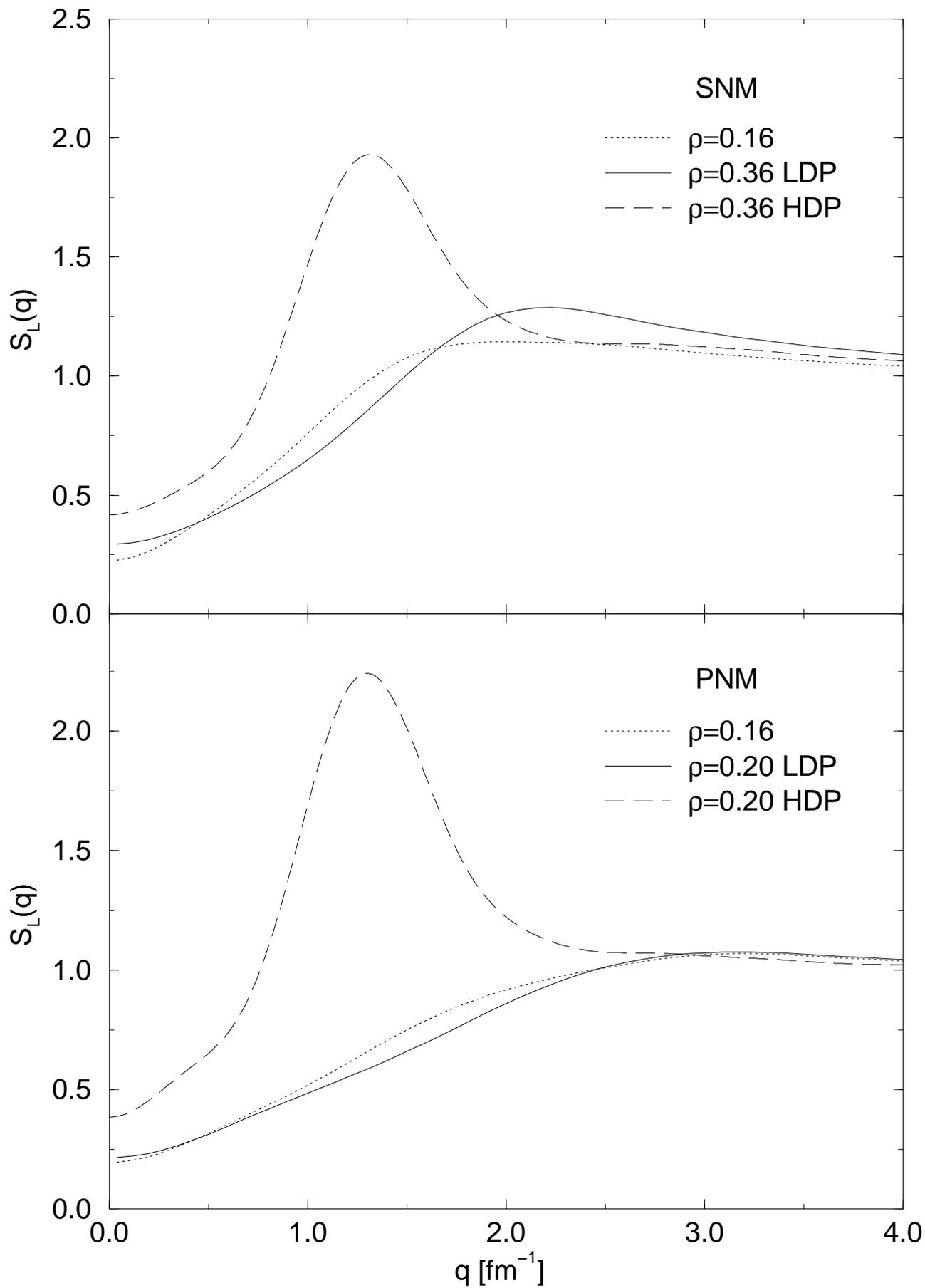
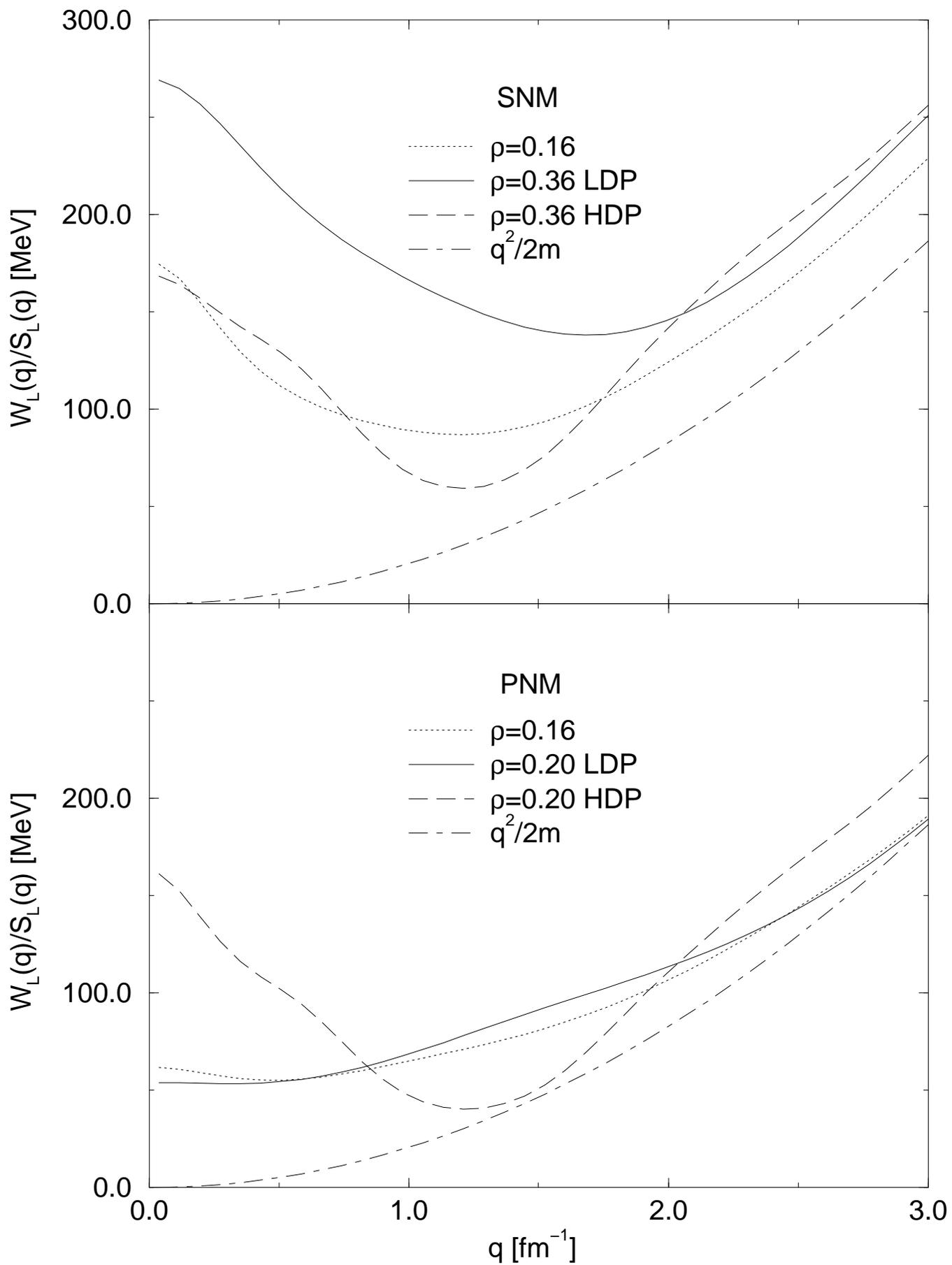
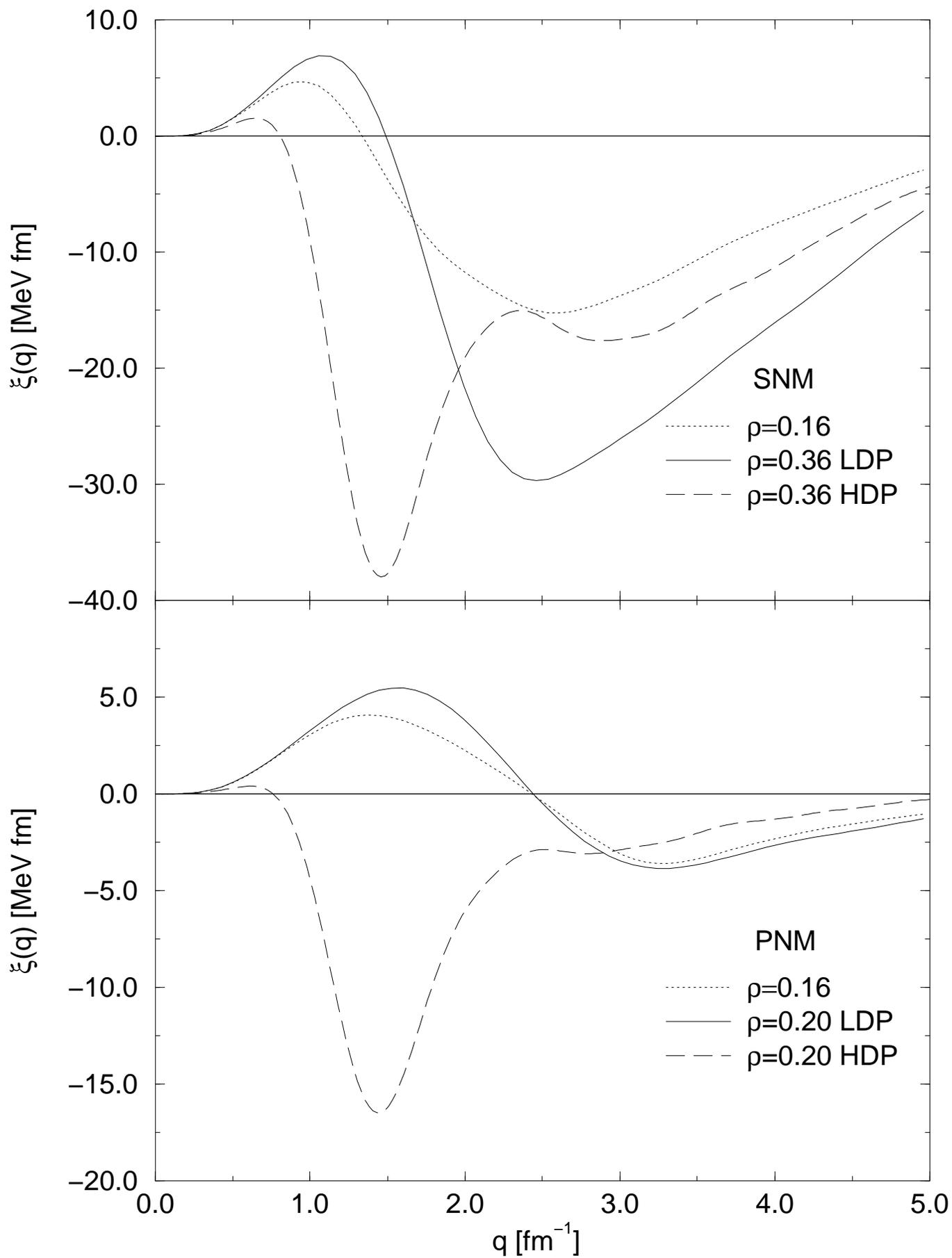
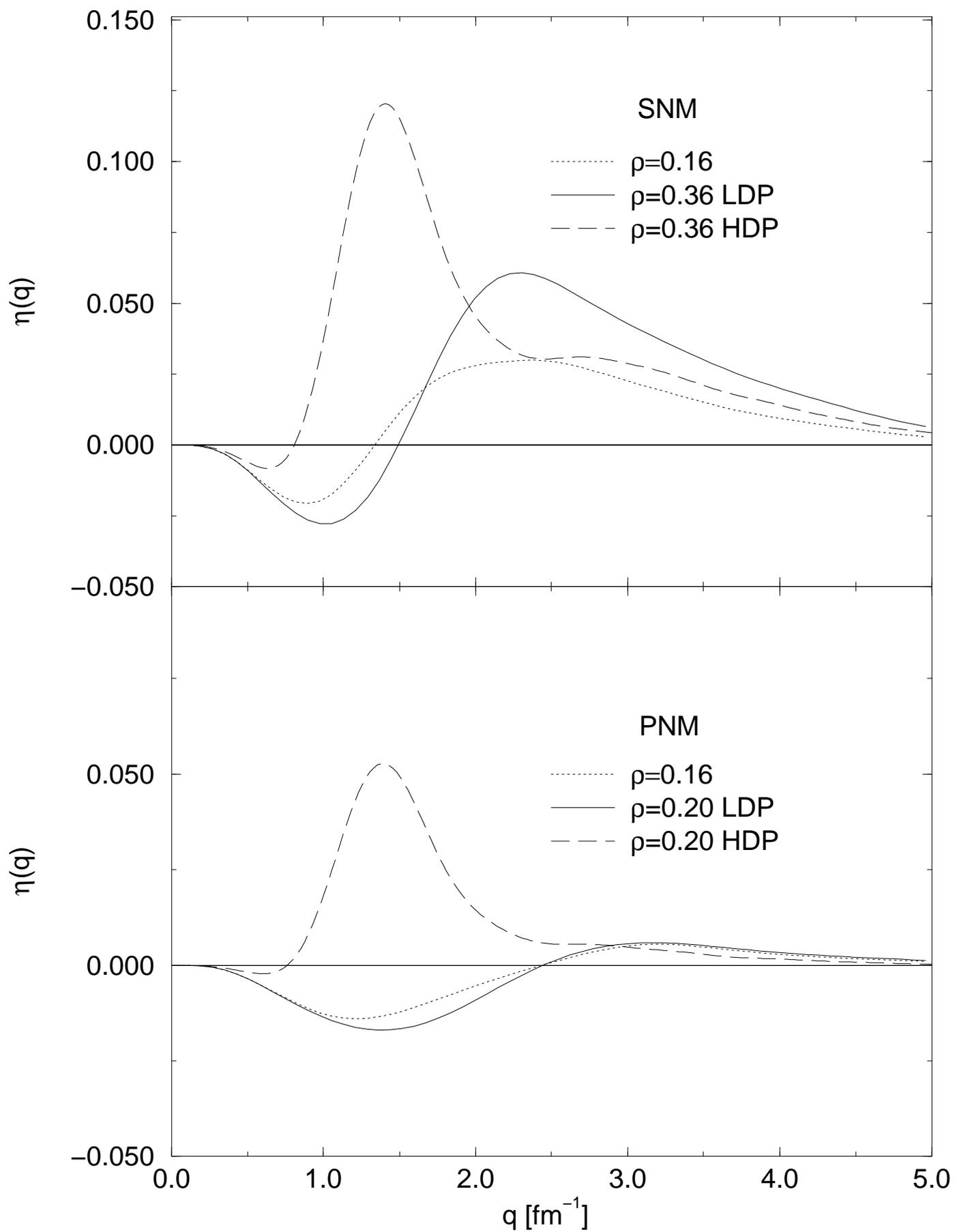
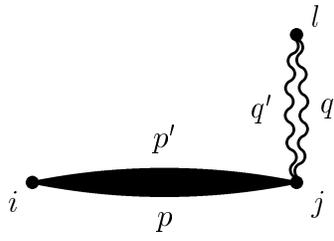
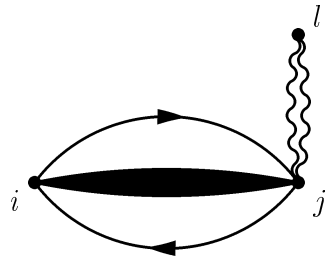
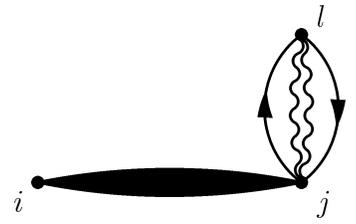
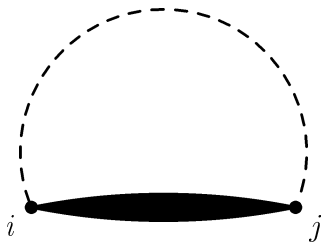
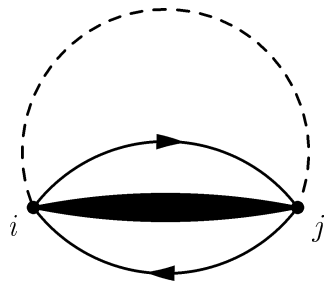
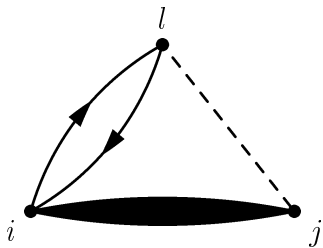
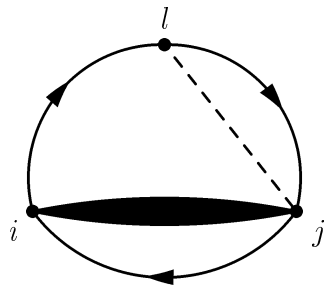
\end{document}